\documentclass[a4paper,11pt]{article}

\usepackage{lmodern}
\usepackage{a4wide}
\usepackage{graphicx} 
\usepackage{verbatim}
\usepackage{amsmath,amssymb,algorithmic,algorithm}
\usepackage{authblk}

\usepackage{amsmath,amssymb,amsthm}

\newtheorem{theorem}{Theorem}[section]
\newtheorem{lemma}{Lemma}[section]
\newtheorem{definition}{Definition}[section]
\newtheorem{remark}[theorem]{Remark}

\newcommand{\x}{\mathbf{x}}

\newcommand{\n}{\mathbf{n}}

\newcommand{\tH}{\tilde{H}}
\newcommand{\hH}{\hat{H}}
\newcommand{\hhH}{\hat{\tilde{H}}}
\renewcommand{\k}{\mathbf{k}}
\newcommand{\p}{\mathbf{p}}
\renewcommand{\d}{\mathrm{d}}

\newcommand{\inZ}{\in \mathbb{Z}}
\newcommand{\R}{\mathbb{R}}
\newcommand{\Z}{\mathbb{Z}}

\newcommand{\erf}{\mathrm{erf}}
\newcommand{\erfc}{\mathrm{erfc}}

\renewcommand{\O}{\mathcal{O}}
\newcommand{\epsi}{\varepsilon}
\newcommand{\kmax}{k_\infty}

\renewcommand{\phi}{\varphi}
\renewcommand{\r}{\mathbf{r}}
\newcommand{\MFT}{\mathrm{MFT}}
\newcommand{\rms}{\text{rms}}
\newcommand{\phiz}{\phi^{F,\k=0}}
\newcommand{\sumn}{\sum_{n=1}^N}
\title{Fast and spectrally accurate Ewald summation for 2-periodic
  electrostatic systems}
\date{September 1, 2011}
\author[1,*]{Dag Lindbo} 
\author[1]{Anna-Karin Tornberg} 
\affil[1]{Numerical Analysis, Royal Inst. of Tech. (KTH), 100 44 Stockholm, Sweden}

\begin{document}
\maketitle
\thispagestyle{empty}

\let\oldthefootnote\thefootnote
\renewcommand{\thefootnote}{\fnsymbol{footnote}}
\footnotetext[1]{To whom correspondence should be addressed. Email:
  dag@kth.se }
\let\thefootnote\oldthefootnote

\begin{abstract}
  A new method for Ewald summation in planar/slablike geometry,
  i.e. systems where periodicity applies in two dimensions and the
  last dimension is ``free'' (2P), is presented. We employ a spectral
  representation in terms of both Fourier series and integrals. This
  allows us to concisely derive both the 2P Ewald sum and a fast
  PME-type method suitable for large-scale computations. The primary
  results are: {\sl (i)} close and illuminating connections between
  the 2P problem and the standard Ewald sum and associated fast
  methods for full periodicity; {\sl (ii)} a fast, {\sl O(N log N)},
  and spectrally accurate PME-type method for the 2P k-space Ewald sum
  that uses vastly less memory than traditional PME methods; {\sl
    (iii)} errors that decouple, such that parameter selection is
  simplified. We give analytical and numerical results to support
  this.
\end{abstract}

\section{Introduction }

\emph{Ewald summation} deals with the task of summing the Coulomb
potential over a set of charged particles that are subject to periodic
boundary conditions. The potential sum itself may be written as
\begin{align}
  \phi(\x) = \sumn \sum_{\p \in \Lambda } \frac{q_n}{\| \x
    - \x_n + f(\p) \|}, \label{eq:pot_sum}
\end{align}
where $\{\x_j, q_j\}, j=1\dots N$, with $\x\in\Omega \subset \R^3$
and $\sum q_j = 0$ (neutrality) , represents the location and charge
of $N$ particles. For simplicity we let $\Omega = [0,
L)^3$. Periodicity is expressed by a translation function $f:\Lambda
\rightarrow \R^3$, and $\Lambda = \Z^d$ denotes indices in a
$d$-dimensional lattice, $d=1,2,3$.  

Complications, which depend on the dimension of periodicity, $d$,
arises because the terms in \eqref{eq:pot_sum} decay $\sim 1/r$. In
fact, a certain amount of ambiguity surrounds direct summation of
\eqref{eq:pot_sum}, see the appropriately named paper by Takemoto
et. al. \cite{Takemoto2003}. 

The present work deals with the accurate (spectrally) and efficient
($N \log N$) computation of the potential sum under two-dimensional
periodicity (the third dimension is ``free''), i.e. when
$\Lambda = \Z^2$, and $f(\p) =[p_1, p_2, 0]$ (see Figure
\ref{fig:2P}). We shall start by briefly surveying the fully periodic
case.

\subsection{3P: Fully extended periodicity }

In the most common situation, periodicity is extended in all three
dimensions, i.e. that $\Lambda = \Z^3$ and $f(\p) = L\p$. This problem
has been thoroughly studied, going back to the eponymous Ewald, who in
1921 \cite{Ewald1921} showed that \eqref{eq:pot_sum} can be computed
by splitting the sum into a rapidly decaying part and a smooth part
which is summed in frequency domain,
\begin{align}
  \phi^{3P}(\x_m) &= \sumn\sum_\p q_n \frac{\erfc(\xi
    \| \x_m - \x_n + L\p \|_2 ) } {\| \x_m - \x_n + L\p \|_2}
  +\nonumber\\
  &\hspace{100pt}+ \frac{4\pi}{L^3}
  \sum_{\k_3\neq 0} \frac{e^{-k^2/4\xi^2}}{k^2} \sumn q_n
  e^{-i\k_3\cdot(\x_m-\x_n)} 
  - \frac{2\xi q_m}{\sqrt{\pi}},
  \label{eq:3dp_ewald_sum}
\end{align}
where $\xi>0$ (which $\phi$ is independent of) is known as the Ewald
parameter, $\k_3\in \{2\pi \n/L : \n \inZ^3\}$, $k=|\k|$, and
the term $(n=m, \p = 0)$ is excluded from the real space sum.

The utility of Ewald summation was greatly enhanced by the development
of $\O(N \log N)$ methods, avoiding the severely limiting $\O(N^2)$
complexity of evaluating \eqref{eq:3dp_ewald_sum} for all $\x_m$. We
denote by \emph{PME} (Particle Mesh Ewald) the well known family of
methods that derive from the pioneering P$^3$M method by Hockney and
Eastwood \cite{Hockney1998}, including major developments such as the
PME method due to Darden et. al. \cite{Darden1993} and the SPME method
by Essmann et. al. \cite{Essmann1995}. The consistency within the PME
family is illustrated in the excellent surveys by Deserno and Holm
\cite{Deserno1998}, and Shan et. al. \cite{Shan2005}. In a survey of
electrostatic calculations in structural biology, an important
application area, Koehl \cite{Koehl2006} points out that the success
of Ewald's method overshadows other methods and that the applications
community has flourished because of this.

\subsection{2P: Planar periodicity }

\begin{figure}
  \centering
  \includegraphics{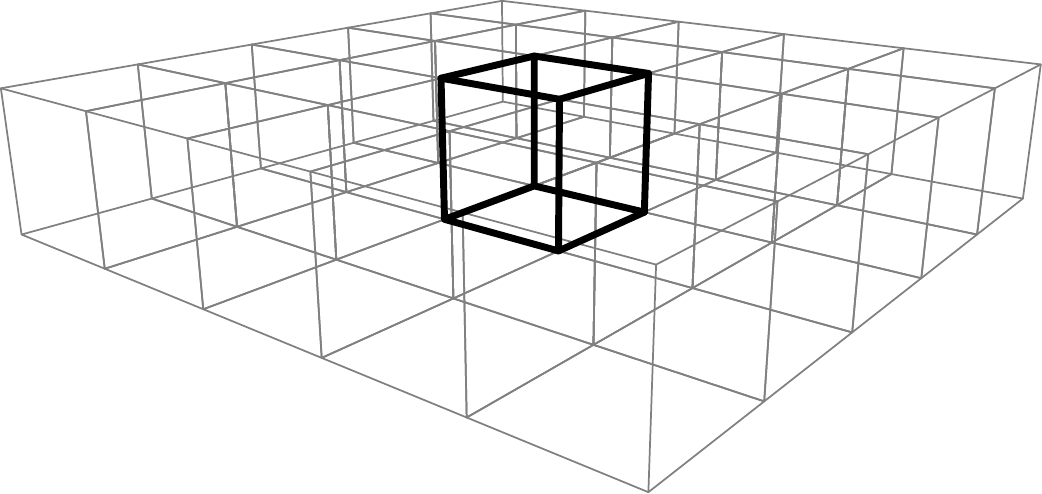} 
  \caption{2P system: Unit cell $\Omega$ (black) repeated infinitely
    in the plane.}
  \label{fig:2P}
\end{figure}

We shall denote the situation when periodicity applies in two
dimensions and the third dimension is free as \emph{planar
  periodicity} or 2P, as illustrated in Figure \ref{fig:2P}. One may
think of a sheet or lamina of charges confined by $z\in[0, L]$ and
infinitely replicated in the $(x,y)$-plane. In the literature this
situation is sometimes referred to as \emph{slab/slablike geometry} or
a \emph{quasi-two-dimensional} system and enjoys a wealth of acronyms,
such as \emph{3D2P1F} (i.e. a three-dimensional system, with two
periodic directions, and one free).

As noted, a satisfactory way to sum the 3P problem came about in the
1920's and work on fast methods took off in the 1990's, based on the
Ewald sum \eqref{eq:3dp_ewald_sum}. In contrast, analysis and methods
for the 2P problem lagged quite far behind, and fast methods have yet
to reach the maturity of their 3P cousins. A summation formula
analogous to \eqref{eq:3dp_ewald_sum} has emerged, but fundamentally
different (i.e. non-Ewald) ideas are also being pursued.

This result, which we shall refer to as the \emph{2P Ewald sum}, was
derived by Grzybowski, Gwozdz and Brodka in \cite{Grzybowski2000}
using lattice sums. Here, the potential sum
\begin{align}
  \phi(\x) = \sumn \sum_{\p \in \Z^2 } \frac{q_n}{\| \x
    - \x_n + \tilde{\p} \|}, \label{eq:2dp_pot_sum}
\end{align}
with $\tilde{\p} = L[p_1, p_2, 0]$, is shown to equal
\begin{align}
  \phi(\x_m) =& \sumn \sum_{\p\in \Z^2}^* q_n \frac{\erfc(\xi
    |\x_m - \x_n +
    \tilde{\p}|)}{|\x_m - \x_n + \tilde{\p}|} + \nonumber \\
  &+\frac{\pi}{L^2}\sumn \sum_{\k\neq 0} \frac{e^{i\k\cdot
      (\r_m-\r_n)}}{k} \bigg[ e^{k(z_m-z_n)}
  \erfc\left(\frac{k}{2\xi} + \xi (z_m-z_n)\right) + \nonumber\\
  &\hspace{170pt} +e^{-k(z_m-z_n)} \erfc\left(\frac{k}{2\xi} - \xi
    (z_m-z_n)\right) \bigg ] + \nonumber \\
  &-\frac{2\sqrt{\pi}}{L^2} \sumn q_n \left( e^{-\xi^2
      (z_m-z_n)^2}/\xi + \sqrt{\pi} (z_m-z_n) \erf(\xi(z_m-z_n))
  \right) -q_m\frac{2\xi}{\sqrt{\pi}}, \label{eq:2dp_ewald_sum}
\end{align}
where $\r \in \R^2$ is the $(x,y)$-component of $\x$, and $\k\in
\{2\pi \n/L : \n \inZ^2\}$.

Their approach follows a classical derivation of the 3P Ewald sum by
de Leeuw et. al. \cite{Leeuw1980}. Interestingly, and as Grzybowski
et. al. point out, the exact same expression can be obtained from much
earlier work by Bertaut \cite{Bertaut1952} and, more recently, by
Heyes et. al. \cite{Heyes1977, Heyes1981, Heyes1981a,
  Heyes1984}. However, it is also attributed to de Leeuw and Perram
\cite{Leeuw1979} by other authors (e.g. \cite{Kawata2001b}). Another
group with a strong claim of independently developing the 2P Ewald
sum is Rhee et. al. \cite{Rhee1989}. Among the foremost in early
developments was Parry \cite{Parry1975,Parry1976}, whose results are
drawn upon by Heyes and others. In the context of the present work, it
is appropriate to highlight Grzybowski et. al. \cite{Grzybowski2000}
as a modern and accessible reference. Irrespective of how one traces
the lineage of \eqref{eq:2dp_ewald_sum}, evaluating it for all $\x_m$
has the dreaded $\O(N^2)$ complexity (with a \emph{very} large
constant) without hinting how a fast method might arise.

There are several alternatives to the 2P Ewald sum
\eqref{eq:2dp_ewald_sum}. An interesting non-Ewald method is known as
\emph{Lekner summation}, due to J. Lekner
\cite{Lekner1989,Lekner1991}, which obtains series that converge
faster than the Ewald sum. The reader is referred to the excellent
survey by Mazars \cite{Mazars2005} for more details, including
comparisons between Lekner and Ewald sums, and appropriate
caveats. Arnold and Holm \cite{Arnold2002a} suggest a ``convergence
factor'' approach -- obtaining a non-Ewald method that goes by the
name \emph{MMM2D}, and is related to the Lekner sum. They, for the
first time, show \emph{a priori} error bounds for the 2P problem.

Another important alternative to \eqref{eq:2dp_ewald_sum} is, somewhat
brazenly, to use the 3P Ewald sum \eqref{eq:3dp_ewald_sum}
instead. The idea here is to extend the unit cell in the $z$-direction,
creating a gap that separates sheets of charged particles (periodicity
in all three directions is implied). Convergence is expected because
the artificial sheets have no net charge. This was investigated by
Spohr \cite{Spohr1997}, where it is indicated, computationally for a
simple system, that \eqref{eq:3dp_ewald_sum} converges to
\eqref{eq:2dp_ewald_sum} as the gap widens. Various methods have been
proposed that introduce correction terms to the 3P sum, such as the
method due to Yeh and Berkowitz \cite{Yeh1999} (see also Crozier
et. al. \cite{Crozier2000}). Present in these references (and in the
works cited therein) are, to a varying extent, additional assumptions
and physically motivated simplifications that do not immediately
generalize. The level of accuracy attained by these methods is seen as
inadequate by present standards.

Moreover, the errors introduced by extending the problem to 3D
periodicity turn out to be quite subtle. In a pair of papers
\cite{Arnold2002,Joannis2002}, Arnold, Holm and de Joannis show that
by formally summing in a planar fashion (rather then spherically, as
is implied in the Ewald sum \eqref{eq:3dp_ewald_sum}) additional terms
emerge. This lets them formulate a correction term which enables high
accuracy and good error control when used in conjunction with their
MMM2D method \cite{Arnold2002a}. They also apply established PME
methods to the extended problem, obtaining a fast method. The work by
Holm et. al. deserves much credit for clarity, appropriate rigor, and
a level of general applicability which is lacking in much of the
preceding work.

There also exists methods that aim to improve the efficiency of
evaluating the 2P Ewald sum \eqref{eq:2dp_ewald_sum}. In a collection
of papers Kawata and collaborators \cite{Kawata2001, Kawata2001a,
  Kawata2002a} propose a method which relies on an integral transform
that was also used by Parry \cite{Parry1975, Parry1976} (see
clarifying correspondence \cite{Mazars2002, Kawata2002}). The same
authors have also proposed a SPME-like method \cite{Kawata2001b} that
relies on the same ideas. However, even the determined reader may
struggle to gain clarity from these sources -- and the practical
accuracy of their methods appears to be low and hard to control. This
is regrettable, as we believe that their basic premises are quite
useful. This shall be elaborated on throughout the present work.

Recent work includes Ewald-related methods due to S. Goedecker and
collaborators, such as the mixed Ewald-finite element method by
Ghasemi et. al. \cite{Ghasemi2007} and related work
\cite{Genovese2007, Neelov2007}.

Of the available methods, the work by Holm et. al. appears to be the
one most widely used. This might well be a consequence of their
proximity to established 3P methods -- which would explain why other
recent work enjoys less attention. For instance, the idea that Ghasemi
et. al. purse \cite{Ghasemi2007} (using a tailored finite element
method in the $z$-direction) adds significantly to the mathematical
and practical complexity of the problem at hand \emph{vis-a-vis} 3P
methods.

We agree with the view promoted by Holm et. al., that there is much
value in having methods for the 2P problem that maintain a close
relationship to the mature 3P methods. However, we believe that
extending the problem to full periodicity, and then laboring
extensively over correction terms to compensate, is a somewhat blunt
approach.

In the present work, we shall use a more subtle approach that avoids
the extension to full periodicity, yet is consistent with the 3P
Ewald framework. It starts from representing functions with planar
periodicity (2P) using both Fourier series (in the periodic
$(x,y)$-directions) and a Fourier integral (in the free
$z$-direction). We show that this admits a natural derivation of the
2P Ewald sum \eqref{eq:2dp_ewald_sum}. Furthermore, we shall see that
an intermediate step in this derivation is a natural starting point
from which a fast, $\O(N\log N)$, and spectrally accurate method can
be developed. We derive this method and motivate it theoretically and
with computational examples.

\section{Ewald summation in planar periodicity} \label{sec:ewald_2dp}

\subsection{Preliminaries} \label{sec:prelim}

Start by defining a set of functions of mixed periodicity (2D periodic
1D free).

\begin{definition}[2P functions]
  Let $V_\Omega$ denote the set of functions $f(x,y,z)$ that are
  periodic in $(x,y)\in\Omega$ and ``free'' in $z\in\R$. Denote
  $V_\Omega \ni f(\r,z), \r=(x,y)\in\Omega$. We shall refer to these
  as functions of mixed periodicity. Functions in $V_\Omega$ have a
  discrete spectrum corresponding to the periodic directions, and a
  continuous spectrum corresponding to $z$. We let $V_\Omega \ni
  f(\r,z) \rightleftharpoons \hat{f}(\k,\kappa)$, $\k\in\{2\pi \n /L : \n
  \in \Z^2 \}, \kappa\in\R$.
\end{definition}

We assume that $f(\r,z)$ and all it's derivatives decay faster than
any inverse power of $z$ in the limit $|z|\rightarrow \infty$, and that
$\int_\Omega |f(\r,z)|^2 \d \r < \infty$ for all $z$. Then $\hat{f}$
exists and represents $f\in V_\Omega$ (with $\Omega=[0, L)^2$):
\begin{align}
  f(\r,z) = \frac{1}{2\pi}\int_\R \sum_{\k}\hat{f}(\k,\kappa) e^{i\k
    \cdot \r} e^{i\kappa z}\d \kappa. \label{eq:k_space_repr}
\end{align}
In fact, in the present work we mostly deal with Gaussians,
$e^{-\alpha r^2}$, i.e. the fixed point of the Fourier transform (a
Schwartz function). As far as spectral properties are concerned, this
is a very strong setting.

We shall need several fundamental results from Fourier analysis,
including Poisson summation, Parseval/Plancherel's formula and the
convolution theorem. Typically, these results are given for either
free-space or periodic functions, see e.g. Pinsky
\cite[Ch. 4]{Pinsky2009} and Vretblad
\cite[pp. 175-181]{Vretblad2003}. For functions in $V_\Omega$ we have
the following:

\begin{lemma}[Poisson summation] 
  \label{lemma:psf_2dp}
  
  Let $f(\x)\in V_\Omega$ have Fourier transform $\hat{f}$, and let
  $L\neq 0$. Then,
  \begin{align*}
    \sum_{\p\in\Z^2} f(\x+\tilde{\p}) = \frac{1}{2 \pi L^2}\int_\R
    \sum_{\k} \hat{f}(\k,\kappa) e^{i\k\cdot\r}
    e^{i \kappa z} \d \kappa,
  \end{align*}
  where $\tilde{\p} = [L\p, 0]$ and $\x=:(\r,z)$.
\end{lemma}

\begin{lemma}[Parseval/Plancherel]

  Let $f(\x)\in V_\Omega$ have Fourier transform $\hat{f}$. Then,
  \begin{align*}
    \int_\R \int_\Omega |f(\r,z)|^2 \d \r \d z = \frac{1}{2\pi}
    \int_\R \sum_{\k} |\hat{f}(\k,\kappa)|^2 \d \kappa.
  \end{align*}
\end{lemma}

\begin{lemma}[Parseval/Plancherel variant]
  \label{lemma:parseval_2dp} 

  Let $f(\x), g(\x) \in V_\Omega$ have Fourier transform $\hat{f}$ and
  $\hat{g}$ respectively. Then,
  \begin{align*}
    \int_{\R} \int_{\Omega} f(\r,z) \overline{g(\r,z)} \d \r \d z =
    \frac{1}{2\pi} \int_{\R} \sum_{\k} \hat{f}(\k,\kappa)
    \overline{\hat{g}(\k,\kappa)} \d \kappa.
  \end{align*}
\end{lemma}

\begin{lemma}[Convolution] 
  \label{lemma:conv}

  The convolution of $f(\x),g(\x) \in V_\Omega$ is defined as
  \begin{align*}
    (f*g)(\r,z) = \int_\R \int_\Omega f(\r-\r',z-z') g(\r',z') \d\r' \d z',
  \end{align*}
  and satisfies 
  \begin{align*}
    h(\r,z) = (f*g)(\r,z) \quad \Longleftrightarrow \quad 
    \hat{h}(\k,\kappa) = \hat{f}(\k,\kappa)\hat{g}(\k,\kappa).
  \end{align*}
\end{lemma}

\subsection{Deriving the 2P Ewald sum } \label{sec:ewald_deriv}

From these definitions and properties we now derive the 2P Ewald sum
\eqref{eq:2dp_ewald_sum} in a way that naturally sets the stage for
our PME-type method (Section \ref{sec:se}). The objective is to
compute
\begin{align*}
  \phi(\x) = \sum_{\p\in \Z^2} \sumn \frac{q_n}{\| \x - \x_n + \tilde{\p}\|},
\end{align*}
where $\tilde{\p} = L[\p, 0], \p\in \Z^2$. The traditional way to
derive the (3P) Ewald sum is to solve the Poisson problem,

\begin{align*}
  -\Delta \phi(\x) = 4\pi\sum_{n=1}^{N} \rho^n(\x), \quad \rho^n(\x) =
  \sum_{\p\in\Z^2} q_n \delta(\x - \x_n + \tilde{\p})
\end{align*}
by introducing a charge screening function, $\gamma(\xi,\x)$, 
\begin{align}
  \rho^n(\x) = \underbrace{\rho^n(\x) - (\rho^n * \gamma)(\x)}_{:=
    \rho^{n,R}(\x)} + \underbrace{(\rho^n * \gamma)(\x)}_{:=
    \rho^{n,F}(\x)}. \label{eq:decomp}
\end{align}
One then builds $\phi$ from 
\begin{align}
  \phi = \sumn (\phi^{n,R} +\phi^{n,F}) \label{eq:poisson_sum}
\end{align}
after solving
\begin{align*}
  \underbrace{-\Delta \phi^{n,R}(\x) = 4\pi\rho^{n,R}(\x)}_{(a)}\quad
  \text{and}\quad \underbrace{-\Delta \phi^{n,F}(\x) =
    4\pi\rho^{n,F}(\x)}_{(b)}.
\end{align*}
The screening function, $\gamma$, is required to go from
$\gamma(\xi,0)=1$ to $\gamma(\xi,\|\x\|\rightarrow \infty) = 0$ with
sufficient regularity and be normalized
$\|\gamma(\xi,\x)\|_{L^2}=1$. The most common choice is a Gaussian,
\begin{align}
  \gamma(\x) = \xi^3 \pi^{-3/2} e^{-\xi^2\|\x\|^2} \quad
  \rightleftharpoons \quad \hat{\gamma}(\k) =
  e^{-k^2/4\xi^2}, \label{eq:gamma}
\end{align}
and the classical Ewald summation result follows from this. Utilizing
this screening function also in the 2P setting, it's a straight
forward computation to solve $(a)$, as it is essentially the same as
in 3P. One arrives at:
\begin{align*}
  \phi^{n,R}(\x) = \sum_{\p\in\Z^2} q_n \frac{\erfc(\xi \|\x - \x_n +
    \tilde{\p}\|)}{\|\x - \x_n + \tilde{\p}\|}, \quad \x\neq\x_n.
\end{align*}
In the limit $\x\rightarrow\x_n$ we wish to remove the
self-interaction, which, under the screening, $\gamma$, has partly
been incorporated into $(b)$,
\begin{align*}
  \lim_{\|\x\|\rightarrow 0} \left( \frac{\erfc(\xi\| \x\|)}{\|\x\|} -
    \frac{1}{\|\x\|}\right) = \lim_{\|\x\|\rightarrow 0}
  -\frac{\erf(\xi\|\x\|)}{\|\x\|} = -\frac{2\xi}{\sqrt{\pi}}.
\end{align*}
Summing, in light of \eqref{eq:poisson_sum}, gives $\sumn
\phi^{n,R}(\x_m) = \phi^R(\x_m) + \phi^S(\x_m)$, with
\begin{align*}
  \phi^R(\x_m) = \phi^R_m &= \sumn \sum_{\p\in\Z^2}^* q_n
  \frac{\erfc(\xi \|\x_m - \x_n +
    \tilde{\p}\|)}{\|\x_m - \x_n + \tilde{\p}\|} \\
  \phi^S_m &= -q_m\frac{2\xi}{\sqrt{\pi}}, 
\end{align*}
where $*$ denotes that the term $\p=0$ is excluded when $n=m$. The
last term, $\phi^S$, is usually referred to as \emph{self interaction}.

The second equation $(b)$ is also treated along the lines of the
classical derivation of the 3P Ewald sum, though mixed periodicity
will play a bigger role here. Additionally, physically motivated
conditions as $z\rightarrow\pm\infty$, consistent with the charge
distribution, have to be satisfied. Returning to
\eqref{eq:k_space_repr}, let
\begin{align*}
  \phi^{n,F}(\r,z) = \frac{1}{2\pi} \int_\R \sum_{\k}
  \widehat{\phi}^{n,F}(\k,\kappa) e^{i\k \cdot (\r-\r_n)} e^{i\kappa
    (z-z_n)}\d \kappa.
\end{align*}
Differentiation gives that
\begin{align}
  -\Delta \phi^{n,F} = \frac{1}{2\pi} \int_\R \sum_{\k}
  (k^2+\kappa^2) \widehat{\phi}^{n,F}(\k,\kappa) e^{i\k \cdot
    \r} e^{i\kappa z}\d \kappa. \label{apa}
\end{align}
On the other hand, using Poisson summation (Lemma
\ref{lemma:psf_2dp}) we get
\begin{align}
  4\pi\rho^{n,F} = 4\pi\sum_{\p\in \Z^2} q_n \gamma(\x-\x_n+\tilde{\p}) & =
  \frac{2 q_n}{L^2} \int_\R \sum_{\k} \hat{\gamma}(\k,\kappa)
  e^{ i \k\cdot(\r-\r_n)} e^{i \kappa (z-z_n)} \d \kappa \nonumber \\
  & = \frac{2 q_n}{L^2} \int_\R \sum_{\k} e^{-(k^2+\kappa^2)/4\xi^2} e^{i
    \k\cdot(\r-\r_n)} e^{ i \kappa (z-z_n)} \d \kappa. \label{banan}
\end{align}
Equating \eqref{apa} and \eqref{banan} gives, for $k^2+\kappa^2>0$,
\begin{align*}
  \widehat{\phi}^{n,F} = \frac{4 \pi q_n}{L^2}
  \frac{e^{-(k^2+\kappa^2)/4\xi^2}}{k^2+\kappa^2},
\end{align*}
so that
\begin{align*}
  \phi^{n,F}(\r,z) = \frac{2 q_n}{L^2}  \int_\R \sum_{\k\neq 0}
  \frac{e^{-(k^2+\kappa^2)/4\xi^2}}{k^2+\kappa^2} e^{i\k \cdot
    (\r-\r_n)} e^{i\kappa (z-z_n)}\d \kappa + \phi^{n,F,\k=0}.
\end{align*}
Up to this point the 2P Ewald derivation has deviated from the
traditional 3P derivation only in the representation formula
\eqref{eq:k_space_repr}. However, there remains to discuss the $\k=0$
term. We write

\begin{align}
  \phi^F(\x_m) = \frac{2}{L^2} \sumn \sum_{\k\neq 0} \int_\R
  q_n \frac{e^{-(k^2+\kappa^2)/4\xi^2}}{k^2+\kappa^2} e^{i\k \cdot
    (\r_m-\r_n)} e^{i\kappa (z_m-z_n)}\d \kappa,\label{eq:tmp1}
\end{align}
so that $\sumn \phi^{n,F}(\r_m,z_m) = \phi^F(\r_m,z_m)+\phiz$.

Before determining the term $\phiz$, one can proceed further with the
integral in \eqref{eq:tmp1}.  Using Erd\'{e}lyi (ed.)  \cite[Ch. 1.4,
(15), p. 15]{Erdelyi1954}, or more the more recent Zwillinger (ed.)
\cite[3.954 (2), p. 504]{Zwillinger2007}, it follows that
\begin{align*}
  \phi^{F}(\x_m) &= \frac{\pi}{L^2}\sumn \sum_{\k\neq 0}
  \frac{e^{i\k\cdot (\r_m-\r_n)}}{k}\bigg[ e^{k(z_m-z_n)}
  \erfc\left(\frac{k}{2\xi} + \xi (z_m-z_n)\right) +\\
  & \hspace{170pt} + e^{-k(z_m-z_n)} \erfc\left(\frac{k}{2\xi} - \xi
    (z_m-z_n)\right) \bigg ].
\end{align*}

Turning to the 2P-specific contribution denoted $\phiz_m$ -- in the 3P
Ewald sum \eqref{eq:3dp_ewald_sum}, the (single) term $\k_3=0$ is
simply dropped due to the condition that $\phi$ integrates to zero,
which is consistent with the charge neutrality constraint, $\sumn
q_n=0$. In the 2-periodic setting the relevant condition takes the
form of a dipole moment with respect to $z$, the non-periodic
direction,
\begin{align*}
  \lim_{z\rightarrow\pm\infty} \phi(\x) = \pm \frac{2\pi}{L^2}
  \sumn q_nz_n.
\end{align*}
The derivation is found in Appendix \ref{app:sing_deriv}, where the
remaining contribution is found to be:
\begin{align*}
  \phi^{F,\k=0}_m = -\frac{2\sqrt{\pi}}{L^2} \sumn q_n \left(
    \frac{1}{\xi}e^{-\xi^2 (z_m-z_n)^2} + \sqrt{\pi} (z_m-z_n)
    \erf(\xi(z_m-z_n)) \right).
\end{align*}

We now have all the terms present in \eqref{eq:2dp_ewald_sum} and the
derivation is complete. To summarize, $\phi$ is computed from
\begin{align*}
  \phi(\x_m) = \phi^R_m + \phi^F_m + \phi^{F,\k=0}_m + \phi^S_m,
\end{align*}
where
\begin{align}
  \phi^R_m &= \phi^R_m = \sumn \sum_{\p\in \Z^2}^* q_n
  \frac{\erfc(\xi |\x_m - \x_n +
    \tilde{\p}|)}{|\x_m - \x_n + \tilde{\p}|}\label{eq:rs_sum}\\
  \phi^F_m &= \frac{2}{L^2} \sumn \sum_{\k\neq 0} \int_\R q_n
  \frac{e^{-(k^2+\kappa^2)/4\xi^2}}{k^2+\kappa^2} e^{i\k \cdot
    (\r_m-\r_n) /L} e^{i\kappa (z_m-z_n)}\d
  \kappa, \quad \text{or as a sum}, \label{eq:fd_integral} \\
  &= \frac{\pi}{L^2}\sumn \sum_{\k\neq 0}
  \frac{e^{i\k\cdot (\r_m-\r_n)}}{k}\bigg[ e^{k(z_m-z_n)}
  \erfc\left(\frac{k}{2\xi} + \xi (z_m-z_n)\right) + \nonumber \\
  & \hspace{170pt} +e^{-k(z_m-z_n)} \erfc\left(\frac{k}{2\xi} - \xi (z_m-z_n)\right)
  \bigg ] \label{eq:fd_sum} \\
  \phi^{F,\k=0}_m &= -
  \frac{2\sqrt{\pi}}{L^2} \sumn q_n \left( e^{-\xi^2
      (z_m-z_n)^2}/\xi + \sqrt{\pi} (z_m-z_n) \erf(\xi(z_m-z_n))
  \right)\label{eq:fd_singular_sum}\\
  \phi^S_m &= -q_m\frac{2\xi}{\sqrt{\pi}} \label{eq:rs_self}.
\end{align}

We shall refer to \eqref{eq:rs_sum} as the \emph{2P real space Ewald
  sum}, to \eqref{eq:fd_sum} as the \emph{2P k-space Ewald sum}. As we
have already pointed out, these expressions have been derived before,
e.g. by Grzybowski et. al. \cite{Grzybowski2000}. However, they
arrive at \eqref{eq:fd_sum} in a completely different manner. For us,
the integral representation of $\phi^F$ \eqref{eq:fd_integral} is the
key result that we shall derive a fast and accurate PME-type method
from. Kawata and Mikami \cite{Kawata2001a} view \eqref{eq:fd_integral}
as a consequence of \eqref{eq:fd_sum}, which is of course valid (the
expressions are equivalent), but runs counter to intuition. The
theoretical foundations set forth in Section \ref{sec:prelim} not only
enable our elementary derivation of
\eqref{eq:rs_sum}-\eqref{eq:rs_self} and the important choice
\eqref{eq:fd_integral} $\vee$ \eqref{eq:fd_sum}, but are also required
as we proceed.

As is well established for the 3P Ewald sum, the infinite sums above
may be truncated (which we elaborate on in Section
\ref{sec:param_choices}). Evaluating \eqref{eq:rs_sum} or
\eqref{eq:fd_sum} $\forall m\in\{1,2,\dots,N\}$ has complexity
$\O(N^2)$ with a \emph{very} large constant (that grows geometrically
as higher accuracy is required). The contribution from the $\k=0$
singularity \eqref{eq:fd_singular_sum} has the same complexity, but
with a smaller constant.

\section{Spectrally accurate fast method for the 2P Ewald sum}
\label{sec:se}

Here we develop a PME-like method with spectral accuracy to compute
the reciprocal space 2P Ewald sums $\phi^F$ \eqref{eq:fd_sum} and
$\phi^{F,\k=0}$ \eqref{eq:fd_singular_sum}. The treatment is
self-contained, but the reader may benefit from being familiar with
our previous paper \cite{Lindbo2011} on the 3P problem.

\subsection{Fast method for $\phi^F$} \label{sec:fast_fd_method}

Consider the computation of the integral form of the 2P $\k$-space
Ewald sum \eqref{eq:fd_integral}:
\begin{align*}
  \phi^F(\x_m) = \frac{2}{L^2} \sum_{\k\neq 0} \int_\R
  \frac{e^{-(k^2+\kappa^2)/4\xi^2}}{k^2+\kappa^2} 
  \sumn q_n e^{i\k\cdot(\r_n
    -\r_m)} e^{i\kappa(z_n-z_m)} \d \kappa.
\end{align*}
We proceed as in \cite{Lindbo2011}, splitting the Gaussian term above
into three parts using a parameter $\eta>0$ (cf. Section
\ref{sec:params}),
\begin{align*}
  \phi^F(\r_m,z_m) =& \frac{2}{L^2} \sum_{\k\neq 0} \int_\R
  \frac{e^{-(1-\eta)(k^2+\kappa^2)/4\xi^2}}{k^2+\kappa^2}
  e^{-i\k\cdot\r_m} e^{-i\kappa z_m} e^{-\eta(k^2+\kappa^2)/8\xi^2} \overline{\widehat{H}(\k,\kappa)} \d \kappa,
\end{align*}
where we have let
\begin{align*}
  \widehat{H}(\k,\kappa) := \sumn q_n
  e^{-\eta(k^2+\kappa^2)/8\xi^2} e^{-i\k\cdot\r_n} e^{-i\kappa z_n}.
\end{align*}
Using the convolution theorem (Lemma \ref{lemma:conv}) and known
transforms one finds
\begin{align}
  H(\r,z) = C \sumn q_n e^{-\beta \|\r-\r_n\|_*^2} e^{-\beta
    (z-z_n)^2}, \quad C = (2\xi^2/\pi\eta)^{3/2}, \beta =
  2\xi^2/\eta, \label{eq:se2p_H_sum}
\end{align}
where $\|\cdot \|_*$ denotes that periodicity is implied (in the
$(x,y)$-plane, \emph{nota bene}). This expression can be efficiently
evaluated on a grid, but to obtain $\widehat{H}(\k,\kappa)$ on a
suitable grid in $\k$-space we must be careful. The computation in the
periodic directions is simple -- just take the FFT -- but in the
$z$-direction one needs to compute the Fourier integral. We refer to
this operation as a \emph{mixed Fourier transform}, $\MFT(\cdot)$, i.e.
\begin{align}
  \hH(\k,\kappa) = \MFT(H(\r,z)), \label{eq:mft}
\end{align}
and clarify this in Section \ref{sec:f_int}. Moving on, let
\begin{align}
  \widehat{\tilde{H}}(\k,\kappa) :=
  \frac{e^{-(1-\eta)(k^2+\kappa^2)/4\xi^2}}{(k^2+\kappa^2)} \widehat{H}(\k,\kappa),
  \label{eq:se2p_H_mult}
\end{align}
so that
\begin{align}
  \phi^F(\r_m,z_m) = \frac{2}{L^2}\sum_{\k\neq 0} \int_\R
  e^{-i\k\cdot\r_m} e^{-i\kappa z_m}
  e^{-\eta(k^2+\kappa^2)/8\xi^2}
  \overline{\widehat{\tilde{H}}(\k,\kappa)} \d
  \kappa. \label{eq:se2p_kint}
\end{align}
To proceed we use Plancherel's Theorem (Lemma
\ref{lemma:parseval_2dp}) with
\begin{align*}
  \hat{f}(\k,\kappa)=e^{-i\k\cdot\r_m} e^{-i\kappa z_m}
  e^{-\eta(k^2+\kappa^2)/8\xi^2},
\end{align*}
and $\hat{g}=\widehat{\tilde{H}}(\k,\kappa)$. Noting that $f$ is a
$\k$-space product between Gaussians and complex exponentials, one may
compute it's inversion, $f(\r,z)$, a as a convolution with
$\delta$-functions. Thus, invoking Lemma \ref{lemma:conv} and known
transforms gives that
\begin{align}
  \phi(\x_m) &= \frac{4\pi}{L^2}\int_\R \int_\Omega \tilde{H}(\r,z)
  \left[C \int_\R \int_\Omega \delta(\r' -\r_m) \delta(z'-z_m)
    e^{-\beta\|\r' -\r_m\|_*^2}e^{-\beta(z'-z_m)^2} \d\r' \d z' \right]
  \d\r \d z
  \nonumber\\
  & = \frac{4\pi}{L^2} \int_\R \int_\Omega \tilde{H}(\r,z) C
  e^{-\beta\|\r -\r_m\|_*^2} e^{-\beta(z-z_m)^2} \d\r \d
  z. \label{eq:se2p_H_int}
\end{align}
Again, $\widehat{\tilde{H}}(\k,\kappa) \longrightarrow
\tilde{H}(\r,z)$ on a grid in real space requires a non-trivial mixed
transform (again, see Section \ref{sec:f_int}), i.e. $\tH(\r,z) =
\MFT^{-1}(\hhH(\k,\kappa))$. The integral \eqref{eq:se2p_H_int} is
evaluated using trapezoidal quadrature (to spectral accuracy). In
contrast to other PME-type methods, the final result
\eqref{eq:se2p_H_int} is an equality -- no approximations have yet
been introduced. Naturally, as the integral is evaluated via
quadrature, and a finite grid for $\x$ is employed, approximations
will enter. We shall see that these errors are well controlled.

To summarize, the algorithm is: $(i)$ compute \eqref{eq:se2p_H_sum},
$(ii)$ take the mixed transform \eqref{eq:mft}, $(iii)$ compute
\eqref{eq:se2p_H_mult}, $(iv)$ take the mixed inverse transform, and
finally $(v)$ compute \eqref{eq:se2p_H_int} at all desired points
$\x_m$.

The key point that we wish to convey with the derivations of Sections
\ref{sec:ewald_deriv} and \ref{sec:fast_fd_method} is \emph{the
  minimal deviation from the treatment of 3P Ewald methods}. In
particular, the fast method presented here is equivalent to
established 3P PME methods with the following exceptions: Gaussians
(rather than e.g.  Cardinal B-splines) are used in the charge
assignment step \eqref{eq:se2p_H_sum}; an integral is evaluated
(rather than interpolation) to get point-values back, in
\eqref{eq:se2p_H_int}; and a Fourier integral replaces the DFT in the
$z$-direction of the transforms.

\subsubsection{Parameterization and modus operandi } \label{sec:params}

\begin{figure}
  % ASY files: gaussian_1, gaussian_2
  \centering
  \includegraphics{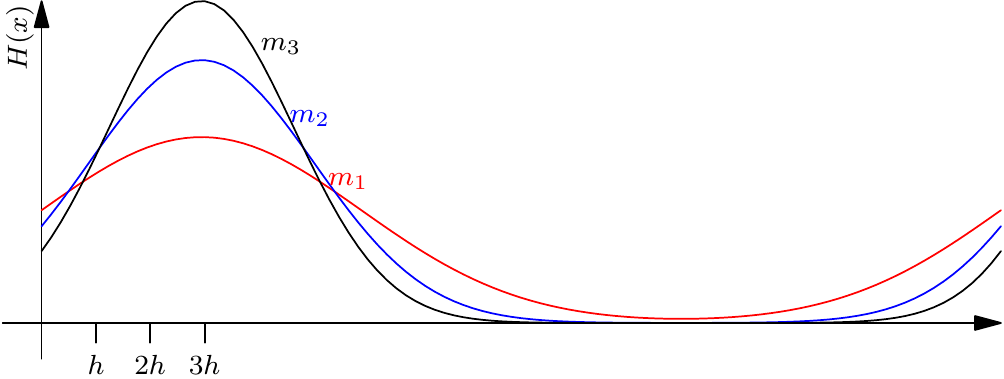}
  \includegraphics{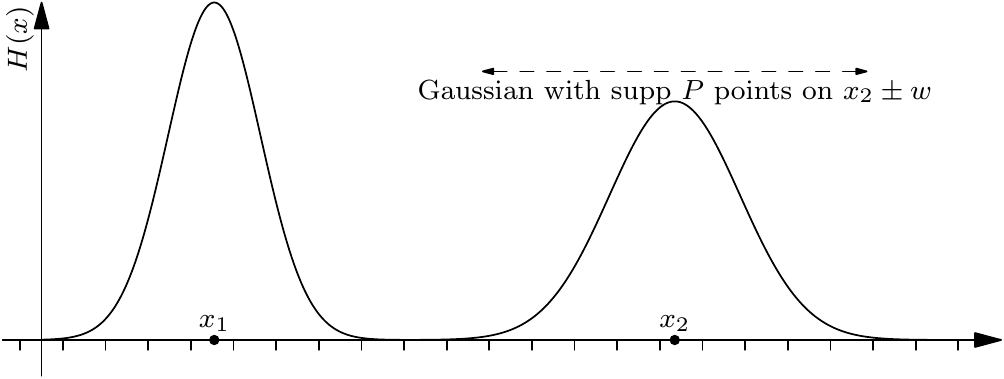}
  \caption{Top: Gaussians with different shape parameters,
    $m_1<m_2<m_3$. Bottom: Gaussian with support on $P$ grid points
    around $\x_j$}
  \label{fig:gaussian}
\end{figure}

There is a free parameter, $\eta$, that can be used to control the
shape of the Gaussian used for the convolutions in
\eqref{eq:se2p_H_sum} and \eqref{eq:se2p_H_int}. We find
(cf. \cite{Lindbo2011}) that a natural choice is to let
\begin{align}
  \eta = \left(\frac{2w\xi}{m}\right)^2, \label{eq:eta_general}
\end{align}
where $w$ represents the half width of a Gaussian and $m$ it's shape
-- see Figure \ref{fig:gaussian}.
Let the domain $\Omega = [0, L)^3$ be discretized with $M$ points in
each direction and let $h=L/M$ denote the grid size. It is implied
throughout that $L$ and $M$ can be different in each direction, and
subscripts will be used when necessary, i.e. $L_z$ and $M_z$.

Gaussians lack compact support, but they are highly localized. It is
natural to truncate them, as is done in the non-uniform FFT
\cite{Greengard2004a,Dutt1993}. We let $P \leq M$ denote the number of
grid points within the support of each Gaussian, as seen in Figure
\ref{fig:gaussian} (bottom). This naturally implies that we take
$w=hP/2$. Furthermore, our analysis shows (cf. Section
\ref{sec:error_analysis}) that the shape parameter, $m$, can be chosen
as $m\sim\sqrt{P}$. This leaves us with a single parameter, $P$.

\begin{remark} \label{rem:M_fixed}
  In contrast to traditional PME methods, we consider the grid size,
  $M$, fixed (determined by the truncation estimates of the Ewald
  sum). The approximation errors added by the fast method are
  controlled by increasing $P$, the number of points within the support
  of each Gaussian.
\end{remark}

\subsection{Computation of Fourier integrals via FFT } 
\label{sec:f_int}

We now make important clarifications regarding the non-trivial mixed
transforms present in the fast method above. Recall that we have two
transforms to compute: $(i)$ from the gridded charge distribution
$H(\r,z)$ to $\hH(\k,\kappa)$, and $(ii)$ from $\hhH(\k,\kappa)$ to
the real-space function $\tH(\r,z)$. Again, $H\in V_\Omega$, and we
think of this mixed periodicity in the following way: $H(\r,z)$ is
periodic in $\r=(x,y)$ and free in $z$ -- hence, $H$ has a transform,
$\hH(\k,\kappa)$, where $\k$ is discrete in the sense that $\k \in
2\pi \Z^2/L$, and $\kappa\in\R$ is a continuous transform variable
corresponding to the non-periodic dimension.

\begin{remark}
  Regardless of how these transforms are computed in practice, it is
  important to remember the underlying mathematics: a 2D discrete
  Fourier transform in (x,y), together with a Fourier integral
  transform in z. The discrete transforms are, of course, computed
  accurately via the FFT, whereas the integral transform raises the
  specter of very large numerical errors. This is one of the defining
  differences between Ewald methods in 2P and 3P.
\end{remark}

Furthermore, we require that the integral transforms can be computed
in the same, $\O(M_z \log M_z)$, complexity as the corresponding DFT
in the 3P method, so as to stay relevant for large-scale
calculations.

We now outline how this is done, based on remarks by Press
et. al. \cite[pp. 692-693]{Press2007}. Let $f(x) \in C^\infty$ decay
sufficiently fast in the the interval $(a,b)$ that it's Fourier
integral transform can be truncated
\begin{align}
  \hat{f}_{a,b}(k) := \int_{a}^{b} f(x) \exp(ikx) \d x,
  \quad \label{eq:ftrans_trunc_1d}
  |\hat{f}(k)-\hat{f}_{a,b}(k)| < \epsi,
\end{align}
for some small $\epsi$. Discretize the interval in real space with $M$
subintervals of size $h = (b-a)/M$. We approximate the integral in a
midpoint fashion,
\begin{align}
  \hat{f}_{a,b}(k) \approx T_M(k) :=& h \sum_{j=1}^{M} f(x_j) \exp(i k
  x_j), \quad \text{with } x_j
  = h(j-1)+a+h/2 = hj +a-h/2, \nonumber \\
  =& h \exp(ik(a-h/2)) \sum_{j=1}^{M} f(x_j) \exp( ikhj
  ). \label{eq:ftrans_quad_meth}
\end{align}
Suppose we want to evaluate $\hat{f}(k)$ on a reciprocal grid
$\frac{2\pi}{L}\{-M/2,\dots, M/2\} \ni k$. Then the remaining sum can
be identified with the discrete Fourier transform, so that
$\hat{f}(k)$ is obtained \emph{on the entire reciprocal grid by a
  single FFT}. 

The task of computing Fourier integral transforms numerically is well
studied in a broader context -- for instance there are the famous
\emph{Filon-type quadratures} (named after L. N. G. Filon who worked
on predecessors to current methods in the 1920's \cite{Filon1928}). A
modern starting point is Iserles and N\"{o}rsett
\cite{Iserles2004,Iserles2004a}, where \emph{matched asymptotic
  expansions} are used to formulate accurate numerical methods for
highly oscillatory integrals. They aim to compute $\hat{f}_{a,b}(k)$
for a single (very large) $k$ with relatively few evaluations of $f$,
under much weaker assumptions on $f$ than we have here. For the
present calculations, we may view the integral as \emph{moderately}
oscillatory. By this we mean that $f$ falls off fast enough that the
maximal characteristic frequency, $(b-a)k_1 \sim M$, of interest is
modest even for very high accuracies (as shall become clear
soon). Furthermore, we need to compute the Fourier integral on all
points $k$ on a reciprocal space grid, and Iserles
\cite[p. 367]{Iserles2004} indicates that an FFT-based method is the
appropriate choice (cf. earlier work by Narasimhan
\cite{Narasimhan1984}).

That said, we return to the midpoint quadrature
\eqref{eq:ftrans_quad_meth}. Press et. al. offer appropriate caution
over this quadrature method: Again, the integral
\eqref{eq:ftrans_trunc_1d} is oscillatory for large $k$, and, since
the maximal $k$ is proportional to $M$, it is not obvious in what
sense $T_M$ converges to $\hat{f}_{a,b}(k)$ as $M$ grows. Of course,
the corresponding inverse Fourier transform may be treated similarly,
and the same caution applies. To investigate the numerical errors
involved we now consider two carefully chosen 1D integrals.

\subsubsection{Fourier integral transform of Gaussian via
  FFT } \label{sec:ftrans_quad_1d}

\begin{figure}
  \centering
  \includegraphics{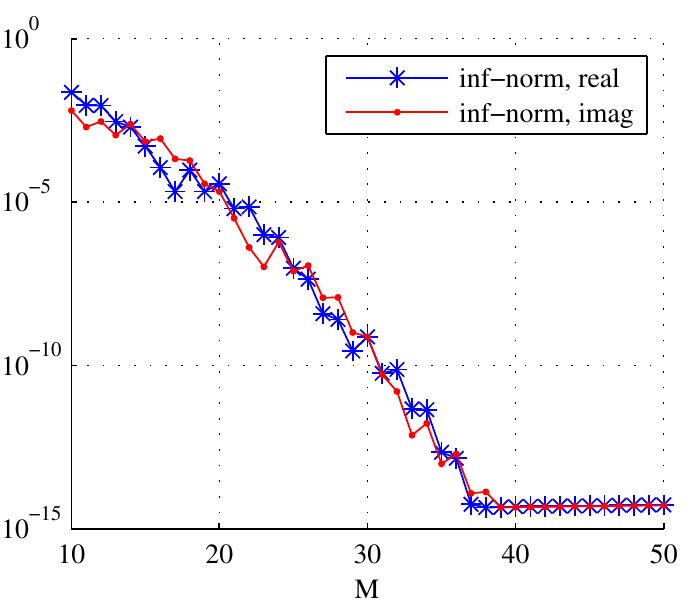} 
  \includegraphics{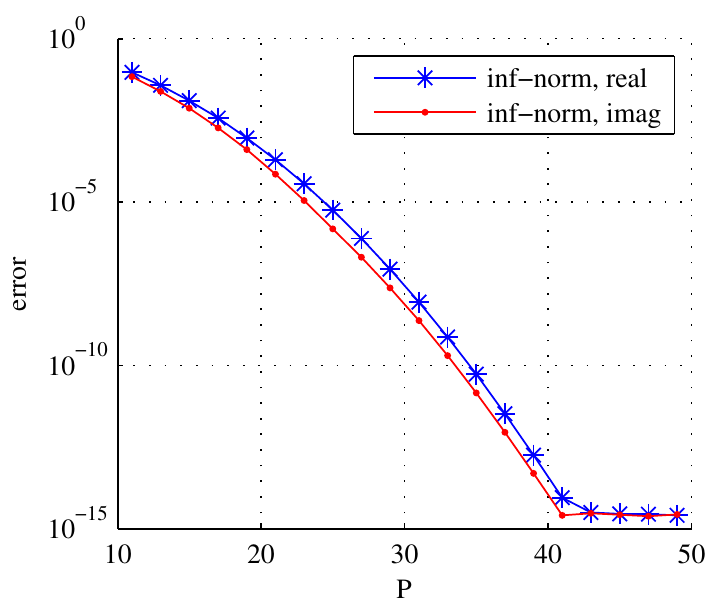} 
  \caption{Left: Convergence of quadrature method
    \eqref{eq:ftrans_quad_meth} for computing Fourier transform of
    Gaussian in 1D \eqref{eq:gauss_transf_1d}. Parameters, $\beta=9$,
    $x'=0.51$, $a=-0.1$, $b=1.1$, chosen to avoid the symmetric cases
    when accuracy comes easier. Right: In the \emph{modus operandi} of
    the proposed method -- convergence of quadrature method, as
    Gaussian support increases, with $M=110$ (fixed), $\xi=8$,
    $a=-0.1$, $b=1.1$, $x'=0.51$, $m=8$.}
  \label{fig:ftrans_quad_I}
\end{figure}

In light of the computations relevant to the present work, we first
restrict ourselves to a Gaussian and its transform,
\begin{align}
  f(x) = \exp(-\beta^2 (x-x')) \quad \rightleftharpoons \quad
  \hat{f}(k) = \sqrt{\pi/\beta^2} \exp(-k^2/(4\beta^2))
  \exp(-ikx'). \label{eq:gauss_transf_1d}
\end{align}
In Figure \ref{fig:ftrans_quad_I} (left) we numerically demonstrate
spectral convergence \eqref{eq:ftrans_quad_meth}, $\|T_M(k)
-\hat{f}(k)\|_\infty \sim \exp(-c(\beta) M^2)$ for some $c>0$ that
naturally depends on $\beta$.

We can go further with the 1D example and introduce the
parameterization, $\beta = 2\xi^2/\eta$ in the transform pair
\eqref{eq:gauss_transf_1d}, and \emph{modus operandi} of the Gaussians
in the fast method above. That is, let the domain $x\in(a, b)$ be
discretized with $M$ points, and let a Gaussian have support on $P$
points around $x=x'$. Considering $M$ fixed and increasing $P$, the
discrete support of the truncated Gaussians, gives the convergence
results in Figure \ref{fig:ftrans_quad_I} (right). Numerically, we
have demonstrated that:

\begin{remark}
  The trivial 1D quadrature method \eqref{eq:ftrans_quad_meth} for the
  Fourier integral transform \eqref{eq:gauss_transf_1d} (with
  $\beta=2\xi^2/\eta$) converges, $\|T_{M,P,m}(k) -\hat{f}(k)\|_\infty
  \sim \exp(-c(m) P^2)$, independently of $M$ and $\xi$.
\end{remark}

Hence, the quadrature required to get into frequency domain has the
important characteristic that approximation errors are controlled by
$P$, the resolution of Gaussians, alone (so that the grid size can be
determined by a truncation estimate for the underlying Ewald
sum). This numerical result is well supported by the analytical error
analysis of our 3P fast Ewald method, cf. \cite{Lindbo2011}. In
essence -- a Gaussian decays fast enough that no numerical
difficulties arise from the oscillatory nature of the Fourier
integral, so the trivial quadrature converges to machine precision
with no need for e.g. \emph{oversampling}.

\subsubsection{Inverse Fourier integral transform of non-Gaussian via
  FFT } \label{sec:iftrans_quad_1d}

Having concluded that the mixed transform of the gridded charge
distribution \eqref{eq:se2p_H_sum} into reciprocal space poses no
particular numerical challenge, we turn to the relevant inverse
transform. With \eqref{eq:se2p_H_mult} in mind we consider
\begin{align*}
  \hat{f}(\kappa) =
  \frac{e^{-(1-\eta)(k_0^2+\kappa^2)/(4\xi^2)}}{k_0^2+\kappa^2} e^{i
    \kappa x'},\quad k_0=2\pi/L, %\label{eq:iftrans_f}
\end{align*}
and the inverse transform
\begin{align}
  f(x) = \frac{1}{2\pi}\int_\R \hat{f}(\kappa)e^{i\kappa x}\d
  \kappa. \label{eq:iftrans_int}
\end{align}
This integral does not offer an obvious closed form as in the previous
example. None the less, for $\eta\leq 1$, the integrand is smooth and
integrable on $\R$. In this particular section we employ
arbitrary-precision integrators from \emph{Mathematica 7}, so that the
FFT-based quadrature method can be evaluated down to the regime of
machine precision.

The quadrature method for the inverse transform is, up to a
normalization, identical to the approximate forward transform
\eqref{eq:ftrans_quad_meth}. Again, we associate the reciprocal space
grid $\frac{2\pi}{L}\{-M/2,\dots, M/2\} \ni k$ with a uniform
staggered grid on the real-space interval $[a, b]$. However, we find
that this reciprocal grid is too coarse. Instead we consider the
family of \emph{oversampled} grids, $\Delta \kappa\{-s_f M/2,\dots,
s_f M/2\}$, $s_f \in \Z^+$, $\Delta \kappa = 2\pi/(s_fL)$. Evidently,
with $s_f=1$ there is no oversampling.

\begin{figure}
  \centering 
  \includegraphics{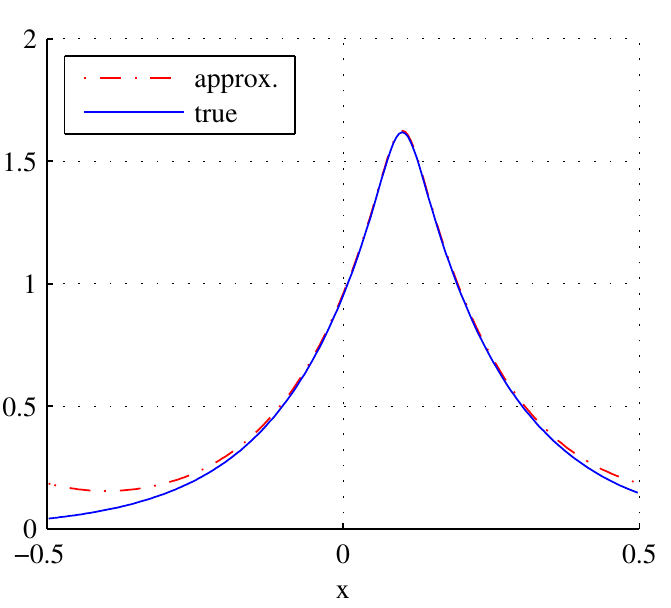} 
  \includegraphics{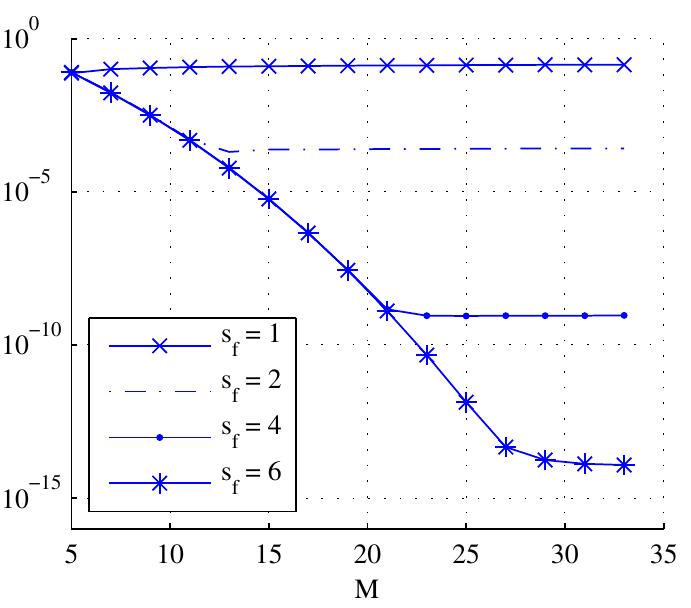}
  \caption{Inverse transform \eqref{eq:iftrans_int}. Left:
    Illustration of non-converging errors when no oversampling
    ($s_f=1$) is used. Right: Convergence of FFT-based quadrature from
    \eqref{eq:gauss_transf_1d}, up to a normalization, to the exact
    transform \eqref{eq:iftrans_int}, $\xi=8$.}
  \label{fig:iftrans_quad_I}
\end{figure}

In Figure \ref{fig:iftrans_quad_I} we present numerical result for a
sequence of grids. Note that without oversampling there are very
visible artefacts of periodicity visible and no convergence. It is
evident that this transform is significantly harder to compute than
the transform of the pure Gaussian (cf. Section
\ref{sec:ftrans_quad_1d}). That said -- oversampling the FFTs by a
small factor (up to six times for double precision accuracy) is well
within the realm of practicality, as we shall return to. It is worth
emphasizing that ``oversampling'' as we defined it here can be given
various other equivalent meanings and monikers, such as ``zero
padding'' and ``$s$ points per wavelength''.

\subsubsection{Extension to functions in $V_\Omega$ }
The results from the 1D analyses generalize to the relevant 3D (or,
rather, 2P) transforms directly. That is, the transform $H(\r,z)
\rightarrow \hH(\k,\kappa)$ is computed via a 3D FFT, where the
``third'' FFT is thought of as a quadrature operation (and appropriate
pre-factors enter). We let $T^{2P}$ denote the immediate extension of
the 1D quadrature scheme \eqref{eq:ftrans_quad_meth} on a grid
$\bar{M}:=[M_{xy}, M_{xy}, M_z]$. One may again pose this computation
in terms of the ``fixed-grid, variable Gaussian support''-setting. To
no surprise, one immediately\footnote{Explicit numerical results for
  the Fourier integrals of 2P functions, analogous to Sections
  \ref{sec:ftrans_quad_1d} and \ref{sec:iftrans_quad_1d} are omitted
  for brevity and concern over repetition. The propositions of spectral
  accuracy of the quadrature in the mixed transforms (and the need to
  oversample the inverse transform) are supported by the numerical
  evaluation of the complete fast method, cf. Section
  \ref{sec:num_res}.\label{fn:repeat}} finds that
\begin{align*}
  \| T^{2P}_{\bar{M},P,m}(\k,\kappa) - \hH(\k,\kappa)\|_\infty \sim
  \exp(-c(m) P^2),
\end{align*}
independently of $\bar{M}$ and $\xi$. Again, this has theoretical
justification in the error estimates of the 3P method
\cite{Lindbo2011}. The inverse transform $\hhH(\k,\kappa)\rightarrow
\tH(\r,z)$, requires over-sampling by at least $s_f=2$ in the third
dimension, as in the 1D example.

The inpatient reader may wonder why we have not precisely defined the
quadrature methods in 2P including the pre-factors for both the
forward and inverse transforms. The reason is that this somewhat
laborious exercise in notation is not needed -- between the forward
and inverse transforms in the fast Ewald method, only a multiplication
\eqref{eq:se2p_H_mult} occurs, so the pre-factors cancel by
linearity. This suggests that we are back to ``just 3D FFTs'' as in
pure 3P Ewald methods. Recall, though, that we are still in the 2P
setting, computing \eqref{eq:rs_sum} - \eqref{eq:rs_self} including
the $\k$-singular contribution \eqref{eq:fd_singular_sum} which we
discuss in Section \ref{sec:fd_singular_sum}. Additionally, we contend
in the next section that an FFT-based quadrature method is efficient
only when the underlying grid function is $C^\infty$ smooth. In
particular, adapting traditional PME-type methods to 2P, as in
\cite{Kawata2001a}, leads to much greater numerical challenges.

\subsubsection{Why not Cardinal B-splines and SPME?  }
\label{sec:cbsp_fquad}

\begin{figure}
  \centering
  \includegraphics{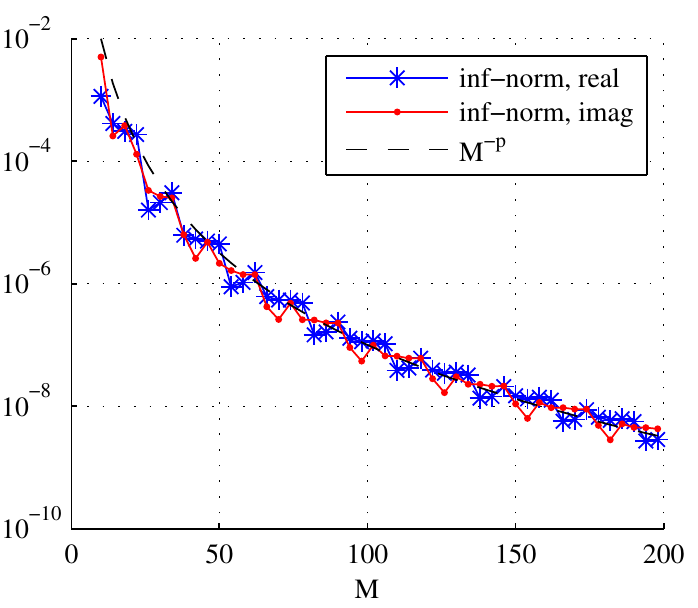}
  \caption{For Cardinal B-spline \eqref{eq:cbsp}, convergence of
    FFT-based quadrature \eqref{eq:ftrans_quad_meth} to exact Fourier
    transform \eqref{eq:cbsp_ftrans}. As expected, the error behaves
    as $M^{-p}$ (dashed line), where $p$ is the order of the B-spline
    . Here, $p=5$.}
  \label{fig:cbsp_ftrans_conv}
\end{figure}

The reader who is familiar with fast Ewald methods may wonder what
role the charge-assignment scheme \eqref{eq:se2p_H_sum} plays for the
computation of the mixed transform. We use Gaussians,
$e^{-\alpha(x-x')^2}$, but that is by no means the only choice. The
\emph{Smooth Particle Mesh Ewald} (SPME) method \cite{Essmann1995},
for instance, uses Cardinal B-splines in the corresponding step. These
are given by
\begin{align}
  F_p(u) = \frac{1}{(p-1)!} \sum_{k=0}^p (-1)^k\frac{p!}{k!(p-k)!}
  (u-k)_+^{p-1}, \quad (x)_+:=\max(x,0) = (x+|x|)/2,
  \label{eq:cbsp}
\end{align}
where $p$ is the order of the spline, and have a known Fourier
transform:
\begin{align}
  \widehat{F_p}(k) = \frac{i(-1+e^{ik})^p}{k^p}. \label{eq:cbsp_ftrans}
\end{align}
Indeed, Kawata and Mikami \cite{Kawata2001a} propose a SPME-like
method for the 2P case that uses this charge assignment function and
their method is also based on the sum/integral
\eqref{eq:fd_integral}. Thus, we ask how the the simple quadrature
method \eqref{eq:ftrans_quad_meth} applied to \eqref{eq:cbsp}
converges to \eqref{eq:cbsp_ftrans}. We note that $F_p \in C^p$ and
$\widehat{F_p} \sim k^{-p}$. In Figure \ref{fig:cbsp_ftrans_conv} we
illustrate the expected convergence, $M^{-p}$, as the number of grid
points in the quadrature method \eqref{eq:ftrans_quad_meth} grows. The
contrast to the convergence in the Gaussian case, Figure
\ref{fig:ftrans_quad_I}, demands more than passing notice.

The loss of regularity by going from Gaussians to Cardinal B-splines
is significant and it lies at the heart of our argument. An FFT-based
quadrature method (as used here and in \cite{Kawata2001a}) loses
\emph{a lot} of accuracy as Gaussians are replace with B-splines in
the integrand. The results given here indicate that hundreds of
grid-points will be needed in the $z$-direction to compute the forward
transform (analogous to the step $H(\r,z) \rightarrow \hH(\k,\kappa)$)
with decent accuracy. Additionally, we saw in Section
\ref{sec:iftrans_quad_1d} that the mixed inverse transform (in our
case $\hhH(\k,\kappa)\rightarrow \tH(\r,z)$) is the main numerical
challenge. We contend that it will be doubly so if Cardinal B-splines,
or any other charge-assignment scheme from 3P PME methods, are
used. The grid sizes and oversampling factors seen in
\cite{Kawata2001, Kawata2001a} seem to support this position.

\subsection{Fast gridding } \label{sec:fgg}

The expressions \eqref{eq:se2p_H_sum} and \eqref{eq:se2p_H_int}
involve computing $N$ exponential functions for each point $\x$ on the
grid. If the grid has $M^3$ points this naively suggests $N M^3$
evaluations of $\exp(\cdot)$, which drops to $N P^3$ with the
truncation from Section \ref{sec:params}. This, as it turns out, is
still many more than are needed if one uses the Gaussian gridding
approach of Greengard and Lee \cite{Greengard2004a}.

The grid-representation of our source distribution
\eqref{eq:se2p_H_sum}, is a sum on the form
\begin{align}
  H(\x) = \left(\frac{\alpha}{\pi}\right)^{3/2} \sumn q_n e^{-\alpha\|\x-\x_n\|^2}.
  \label{eq:fgg_H_sum}
\end{align}
For clarity here we shall suppose that the Gaussians are not
truncated. The key observation is that we wish to evaluate $H(\x)$ on
an equidistant grid, i.e. $\x = [ih,jh,kh]$, where $(i,j,k)$
are integer index triplets in the range $0,1,\dots,M-1$. To see how we
can reduce the number of computations of $\exp(\cdot)$, take the
analogous 1D Gaussian,
\begin{align}
  e^{-\alpha(x-x_n)^2} =  e^{-\alpha(ih-x_n)^2} = e^{-\alpha( (ih)^2-2ih x_n +x_n^2)}
  \nonumber\\
  = \underbrace{e^{ -\alpha(ih)^2 }}_{(a)}
  \big(\underbrace{e^{2\alpha h x_n}}_{(b)}\big)^i
  \underbrace{e^{-\alpha x_n^2}}_{(c)}. \label{eq:fgg_1d}
\end{align}
Note that the term $(a)$ is independent of $x_n$, so those $M$
evaluations of $\exp(\cdot)$ are done once, stored and reused for each
of the $N$ sources $x_n$. The terms $(b)$ and $(c)$ each incur one
$\exp(\cdot)$ for each $x_n$. The same procedure is then applied for
$e^{-\alpha(y-y_n)^2}$ and ditto for $z$. For full algorithms,
additional details and \emph{important} remarks, we refer to
\cite{Lindbo2011}. The bottom line is that, rather than having to
compute $NP^3$ exponentials, the gridding step requires $P^3 + 4N$
exponentials and $\O(NP^3)$ multiplications. This translates to a
significant performance gain in practice.

\subsection{Fast method for $\phi^{F,\k=0}$ } 
\label{sec:fd_singular_sum} 

Turning now to an efficient and accurate method for evaluating the
singular part of the reciprocal space 2P Ewald sum: The computation
of \eqref{eq:fd_singular_sum},
\begin{align}
  \phi^{F,\k=0}_m = \phi^{F,\k=0}(z_m) = - \frac{2\sqrt{\pi}}{L^2}
  \sumn q_n f(z_m-z_n)
  \\
  f(z) = e^{-\xi^2 z^2}/\xi + \sqrt{\pi} z \erf(\xi z), \label{eq:k0_fast_f}
\end{align}
is much less complex than the fast computation of
\eqref{eq:fd_integral} -- it is a finite sum over terms that only
depend on $z$. The obvious approach to avoid $\O(N^2)$ complexity is
an appropriate interpolation method (sometimes imprecisely referred to
as \emph{table lookup}), and the natural choice is to use Chebyshev
polynomials. This method is close to optimal in $\infty$-norm and
cheap to compute (even though there is no periodicity in
$\phi^{F\,\k=0}(z)$). More precisely, we have $z_n \in [0, L_z]$, and
let $p_k, k=1,2,\dots,M_T \ll N$, be the set \emph{Chebyshev-Gauss}
points $\cos(\pi(2k-1)/(2M_T))$ scaled to the interval $[0, L_z]$. We
expand $\phi^{F,\k=0}$ in terms of Chebyshev polynomials
\begin{align*}
  \phi^{F,\k=0}(z) \approx T(z) = \sum_{j=1}^{M_T} c_j T^{[0,L_z]}_j(z),
  \quad \phi^{F,\k=0}(p_k) = T(p_k),
\end{align*}
where $T^{[0,L_z]}_j(z)$ is the $j$:th Chebyshev polynomial scaled to
the relevant interval. The coefficients $c_j$ are easily computed
after evaluating $\phi^{F,\k=0}(p_k)$, and $M_T$ is in essence an
accuracy parameter -- so the complexity of this task is $\O(N)$. We
are dealing with interpolation in 1D, so the computational resources
involved are entirely trivial\footnote{As a point of reference, with
  $N=10^6$ particles and $M_T=40$ Chebyshev polynomials it takes
  roughly one second to evaluate $\phi^{F,\k=0}(z_m)$ at all points
  $z_m$, $m=1,\dots,10^6$.}. Note that the well-known Clenshaw formula
should, for reasons of numerical stability, be used when evaluating
the orthogonal basis $\sum c_j T_j(z)$.

\begin{figure}
  \centering
  \includegraphics{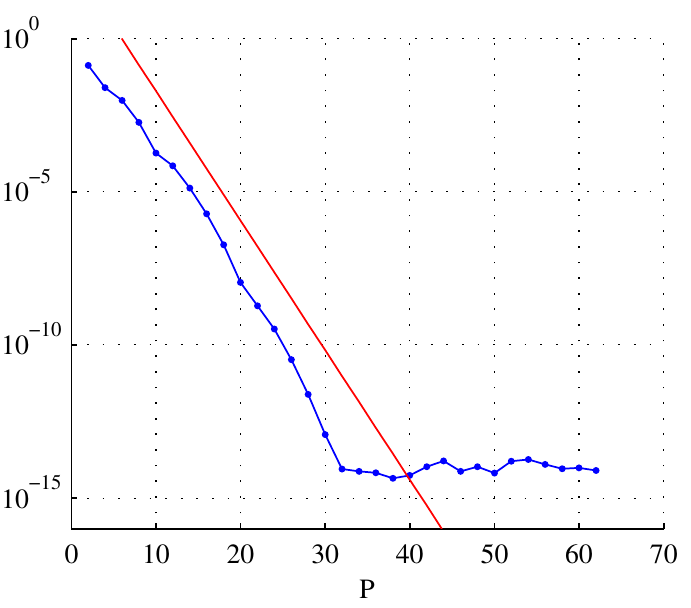}
  \includegraphics{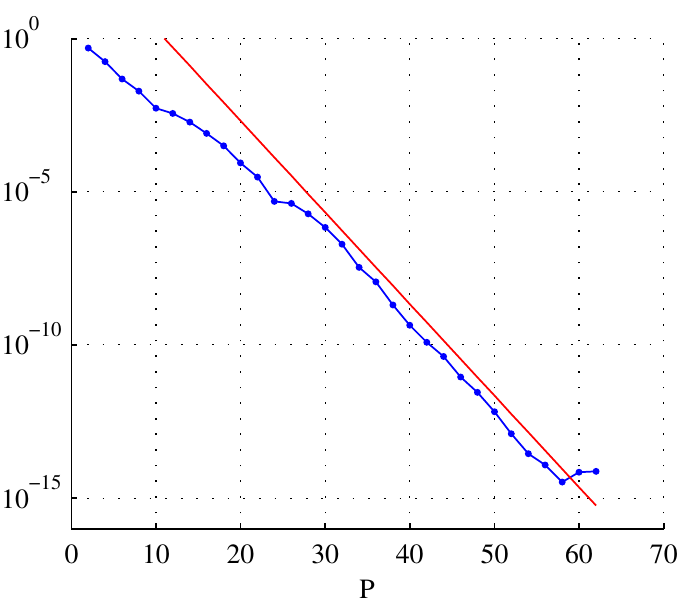}
  \caption{Error in $\infty$-norm when computing $\phi^{F,\k=0}$ with
    Chebyshev interpolation, with $N=10$, together with ``practical''
    error estimate \eqref{eq:interp_est}. Left: $\xi=5$. Right: $\xi =
    10$.}
  \label{fig:k0_fast_convg}
\end{figure}

Very strong error bounds exist for Chebyshev interpolation (see
e.g. the classical references Rivlin \cite{Rivlin1990,Rivlin2010}),
such as
\begin{align*}
  e_M := \max_{z\in [0, L_z]}\big|f(z) - T^f(z)\big| \leq
  \frac{f^{(M_T)}(c)}{2^{M_T-1} M_T!}, 
\end{align*}
for some $c\in [0, L_z]$, where $T^f$ is the interpolant of order
$M_T$ to $f$ \eqref{eq:k0_fast_f}. It is evident that $f$ possesses
$M_T$ derivatives. However, each differentiation yields roughly a
factor $\xi^2$, so the interpolation error will ultimately depend on
$\xi$. This suggests an interpolation estimate of the form $e_M
\approx C\xi^{M_T} 2^{-(M_T-1)}/M_T!$, but that turns out to be
impractical and inaccurate due to the very large quantities
involved. Instead, we find that
\begin{align}
  e_M \approx \xi^3 2^{-c(M_T-1)},\quad c=\pi \xi^{-1/2}, \label{eq:interp_est}
\end{align}
provides a useful form. Again, one may treat error estimation here
with some laxity, as the performance penalties associated with being
cautious (i.e. taking $M_T$ needlessly large) are \emph{very}
small. We give brief numerical results in Figure
\ref{fig:k0_fast_convg}.

In their method, Kawata and Mikami \cite{Kawata2001a} propose a
similar approach based on B-splines. These, of course, have polynomial
accuracy order. A small numerical experiment (omitted) indicates that
the previous remarks about computational triviality then fail to apply
(at least in the broad accuracy regime considered).

\subsection{Error analysis} \label{sec:error_analysis}

We now gather strands of numerical and theoretical results into an
aggregate view of the numerical properties of the proposed
method. This serves the dual purposes of putting previous statements
of accuracy on secure theoretical foundations, and providing useful
guidance for the often intricate task of parameter selection. As
previously alluded to, we start from the classification of numerical
errors into two categories: errors that stem from the underlying Ewald
sum \eqref{eq:2dp_ewald_sum} and errors that stem from the fast method
of the present section.

\subsubsection{Truncation estimates for Ewald sums}
\label{sec:truncation}

Turning to the 2P Ewald sums for $\phi^R$ \eqref{eq:rs_sum} and
$\phi^{F}$ \eqref{eq:fd_sum}, we note that both sums are infinite
but rapidly converging. The real space sum \eqref{eq:rs_sum} is
unchanged \emph{vis-a-vis} 3P, and has been thoroughly studied in
that context. The reader may already be familiar with the famous
error estimates by Kolafa and Perram \cite{Kolafa1992}, which
suggest that the truncation error committed by letting $\|\x\|<r_c$
may be estimated by 
\begin{align}
  e^R_{\rms}(r_c,\xi) \approx \sqrt{\frac{Q}{2L^3}} (\xi r_c)^{-2} e^{-r_c^2
    \xi^2},
  \label{eq:real_trunc_est}
\end{align}
where $Q:=\sumn q_n^2$ and the RMS norm is defined as
$e_{\rms}:=\sqrt{N^{-1}\sumn(\phi_n-\phi^*(\x_n))^2}$. If
particles are statistically correlated, i.e not randomly scattered,
an $\infty$-norm measure may be more appropriate, see e.g. Strain
\cite{Strain1992}.

Correspondingly, for the 3P $\k$-space Ewald sum -- truncated at finite
number of modes, $\kmax\in\Z^+$, i.e. $\|\k\|\leq 2\pi\kmax/L$ -- Kolafa \&
Perram \cite{Kolafa1992} suggest that
\begin{align}
  e^F_{\rms}(\kmax,\xi) \approx \xi \pi^{-2} \kmax^{-3/2}\sqrt{Q} \exp
  \left( -\left( \frac{\pi\kmax}{\xi L}\right)^2\right).
  \label{eq:fd_trunc_est}
\end{align}

\begin{figure}
  \centering
  \includegraphics{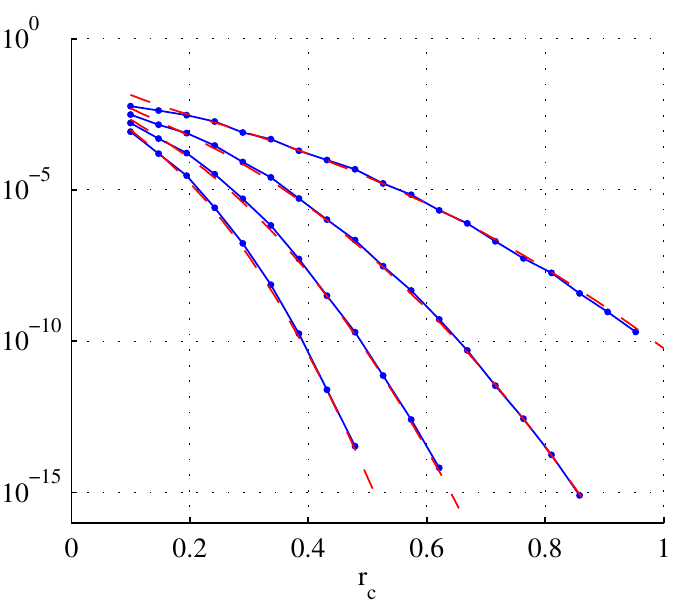} 
  \includegraphics{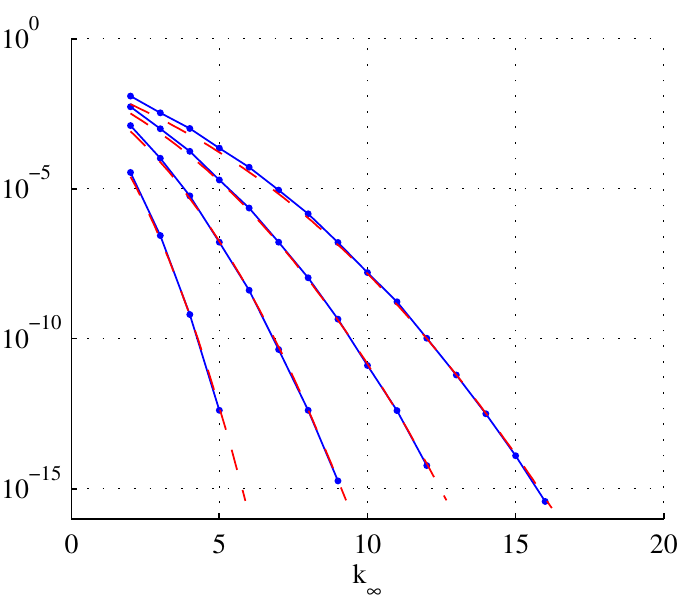}
  \caption{Left: Convergence of real-space sum \eqref{eq:rs_sum} as
    function of truncation radius, $\|\x\|<r_c$, for (right to left)
    $\xi=4,6,8,10$ and the error estimate \eqref{eq:real_trunc_est} in
    dashed. Right: convergence of $\k$-space sum \eqref{eq:fd_sum} as
    a function of the truncation, $\|\k\|\leq 2\pi\kmax/L$, for (left to
    right) $\xi=4,6,8,10$ and the error estimate
    \eqref{eq:fd_trunc_est} in dashed.}
  \label{fig:trunc_est}
\end{figure}

The feasibility of \eqref{eq:fd_trunc_est} as a 2P estimate may come
as a surprise, as it arose from analysis of the 3P sum. We contend
that this is quite natural -- roughly speaking, each dimension has to
converge. Figure \ref{fig:trunc_est} shows numerically that
\eqref{eq:real_trunc_est} and \eqref{eq:fd_trunc_est} capture the
behavior of the truncation error with striking agreement.

\subsubsection{Approximation errors } \label{sec:approx_err}

The second family of errors are those that stem from the fast method,
described in the preceding sections, notably the error due to the
quadrature used to evaluate \eqref{eq:se2p_H_int}. An extensive
treatment is given in \cite{Lindbo2011}, where we prove the following
theorem:

\begin{theorem}[Error estimate] \label{thm:err_est} Given $\xi>0$,
  $h>0$ and an odd integer $P>0$, let $w=hP/2$, and define $\eta$
  according to \eqref{eq:eta_general}.  The error incurred in
  evaluating \eqref{eq:se2p_H_int} by truncating the Gaussian at
  $\|\x-\x_m\|=w$ and applying the trapezoidal rule $T_p$ can be
  estimated by
  \begin{align}
    | \phi - T_P | \leq C\left( e^{-\pi^2 P^2/(2 m^2)} +
      \erfc\left(m/\sqrt{2}\right)\right). \label{eq:quad_err}
  \end{align}
\end{theorem}
From this we surmise an appealing choice of the shape parameter, $m(P)
\approx \sqrt{\pi P}$, which then yields a quadrature error estimate
\begin{align}
  E_Q(P) := | \phi^F - T_P | \approx C e^{-\pi P/2}, \label{eq:quad_est}
\end{align}
to be verified in Section \ref{sec:num_res}. Two other errors emerge,
specific to the 2P method: First, there is an interpolation error
from the fast method for $\phi^{F,\k=0}$, as we investigated in
Section \ref{sec:fd_singular_sum}. Secondly, there is the need to
oversample the inverse transform when computing $\tH(\r,z)$, which we
devoted Section \ref{sec:iftrans_quad_1d} to. We view the oversampling
guidelines from Figure \ref{fig:iftrans_quad_I} as generally
applicable, and content ourselves with that.

\subsubsection{Choosing parameters } \label{sec:param_choices}

There are several parameters present in all (fast) Ewald methods and
they should be chosen with two goals in mind: balancing the work
between the real- and $\k$-space sums (by choosing $\xi$), and
attaining a desired accuracy (selecting e.g. an appropriate PME
grid $M$). The first concern is inherently implementation-dependent
and work-balance will depend strongly on $N$, the number of
charges. \emph{Ipso facto}, there can not exist an optimal parameter
set of broad applicability, and there is no generally accepted tuning
method.

The second concern, assuring that the end-result satisfies a desired
accuracy is also an open question (and likely to remain that way). Two
approaches stand out in the literature: using an optimization
technique, and relying on $a priori$ error estimates. Among the
advocates of ``optimization''\footnote{A more correct description of
  the parameter optimization problem may be ``scanning'', and should
  generally not be confused with numerical optimization techniques
  (such as gradient-based methods).} are Kawata
et. al. \cite{Kawata2001b} and Ghasemi et. al. \cite{Ghasemi2007}. The
former (cf.  \cite[Tab. 2]{Kawata2001b}, where eight parameters are
determined) suggest a high degree of irregularity in the parameter set
(which is either incorrect \emph{per se} or grounds for concern over
the numerics involved). In the latter work, ``Pareto frontier
optimization'' is used (on a set of five parameters) for a specific
crystalline system. In both cases, it is implied that a similar
investigation should be performed whenever a new system is under
consideration -- but the optimization technique depends on a
sufficiently accurate reference solution being computable by the
underlying Ewald sum, which naturally restricts the method to small
$N$.

On the other hand, relying on \emph{a priori} error analysis alone has
to confront a complicated mix of numerical errors. Whereas the picture
is clear for pure 3P Ewald summation \eqref{eq:3dp_ewald_sum}, as we
discuss in Section \ref{sec:truncation}, fast methods often pose stiff
challenges to error estimation. In their survey of fast 3P methods,
Deserno and Holm \cite{Deserno1998} sketch the parameter space and
remind us that many numerical errors are interdependent. Methods for
the 2P situation are less mature and fewer error estimates have been
established. A notable exception is the MMM2D method by Holm
et. al. \cite{Arnold2002a, Arnold2002}, which enjoys sharp error
estimates that are suitable for parameter selection. In
\cite{Joannis2002}, they provide analysis for the case when the 3P
Ewald sum is used for 2P systems, though results are absent for the
PME-accelerated case.

Our view is that error analysis should be the primary focus, but a
certain amount of experimentation is a worthy complement. A particular
goal is to have errors that decouple, so that parameters can be chosen
in sequence and numerical experiments can treat one parameter at a
time. This is by no means simple -- established PME methods for the 3P
Ewald sum have approximation and truncation errors in a tangle after
doing charge-assignment by e.g. B-splines, as we elaborate on in
\cite{Lindbo2011} -- but in the present work decoupling is
achieved. Furthermore, to be useful, error estimates need to be sharp
and simple enough that they are ``solvable'' for a desired parameter.

As a sequence of considerations we suggest
\begin{enumerate}
\item Determine a truncation radius, $r_c$, such that the real space
  sum is cheap to compute (cf. \emph{neighbor list} methods
  \cite{Allen1989} or as summarized in \cite{Lindbo2011}).
\item Select Ewald parameter, $\xi$, such that the real space sum has
  converged to within a given tolerance, $\epsi$, at $\|\r\|<r_c$ by invoking
  e.g. \eqref{eq:real_trunc_est},
  \begin{align}
    \xi = \frac{1}{r_c}\sqrt{W\left(\frac{1}{\epsilon
        }\sqrt{\frac{Q}{2 L^3}}\right)}, \label{eq:xi_est}
  \end{align}
  where $W(\cdot)$ is the \emph{Lambert W-function} (also known as the
  \emph{product logarithm}, it is among the ``special functions''
  provided in e.g. Matlab and Mathematica, defined as the
  inverse of $f(W) = W e^{W}$ \cite{Corless1996}).
\item Then determine the truncation, $k_\infty$, of the $\k$-space
  Ewald sum \eqref{eq:2dp_ewald_sum}, such that the same tolerance is
  met, from \eqref{eq:fd_trunc_est}:
  \begin{align}
    k_\infty > 
    \frac{\sqrt{3} L \xi}{2 \pi }\sqrt{W\left(\frac{4 Q^{2/3}}{3 \pi
          ^{2/3} L^2 \xi ^{2/3} \epsilon ^{4/3}}\right)}.
    \label{eq:k_M_est}
  \end{align}
  This gives the grid size to be employed in the spectral PME method:
  $M = 2 k_\infty$.
\item The quadrature error estimate \eqref{eq:quad_est}, implies the
  number of points within the support of each Gaussian:
  \begin{align*}
    P > -\frac{2L\log(\epsi/C)}{\pi}.
  \end{align*}
  Note that one may get $P>M$ (if $\xi$ small), in which case one has
  to increase the grid size, i.e. $M=\max(P,2 k_\infty)$. The
  constant, $C$, here (from the error estimate \eqref{eq:quad_est})
  does not depend on $\xi$, so $P$ is perhaps most conveniently
  identified from a basic convergence test, e.g. Figure
  \ref{fig:se2p_pme_acc} where the estimate is plotted with $C=10$.

\item Select a oversampling factor, $s_f$, for the reciprocal space
  calculations of the fast method, as discussed in Section
  \ref{sec:f_int}.
\item Finally, determine the Chebyshev grid (cf. Section
  \ref{sec:fd_singular_sum}) for computing $\phi_m^{\k=0}$
  \eqref{eq:fd_singular_sum}.
\end{enumerate}

If this sequence of steps seems laborious, it might be worth pointing
out that having a spectrally accurate method at hand makes it cheap to
err on the side of caution -- the parameter with the greatest impact
on run-time is $P$. Again, it's a sequential process -- rather than a
non-linear optimization problem.

\section{Numerical Evaluation} 

\subsection{Accuracy of spectral PME method } \label{sec:num_res} 

\begin{figure}
  \centering
  \includegraphics{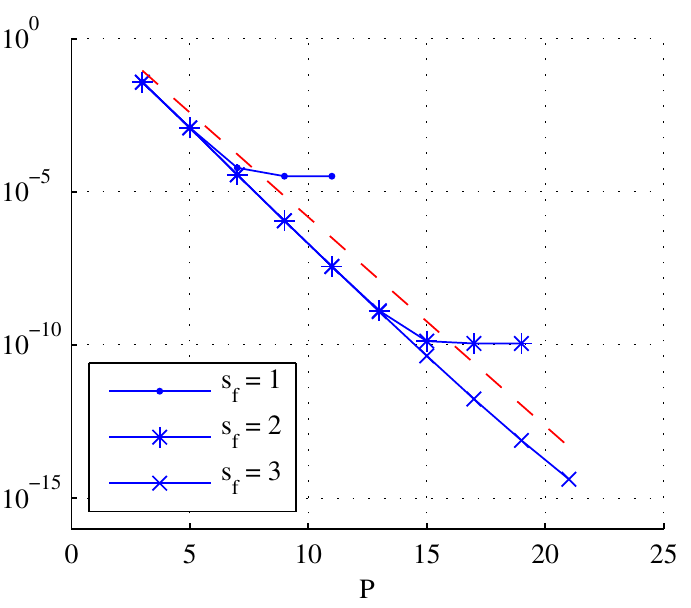}
  \includegraphics{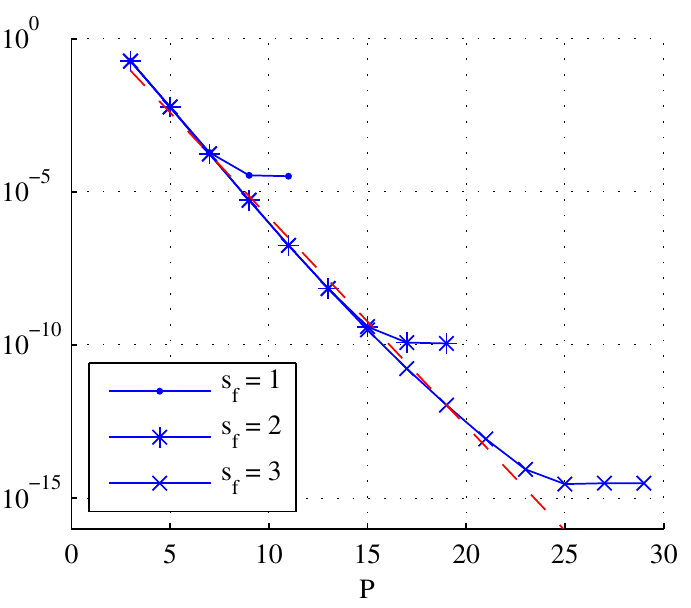}
  \caption{Convergence of SE2P method to Ewald sum, in norm
    \eqref{eq:err}, for various oversampling factors, with quadrature
    error estimate \eqref{eq:quad_est} as dashed line. The
    computational domain is $[0,1)^2\times[-1/2, 3/2]$, so $M_z=2 s_f
    M$. Left: $\xi=4$, $M=21$. Right: $\xi=12$, $M=50$.}
  \label{fig:se2p_pme_acc}
\end{figure}

In Section \ref{sec:error_analysis} we showed that convergence of the
order $e^{-\alpha P}$ is to be expected (we have taken $m =
0.91\sqrt{\pi P}$). To test this we consider small systems, so that
the 2P Ewald sum \eqref{eq:2dp_ewald_sum} can be accurately computed
as a reference, denoted $\phi^*$, and measure errors in the following
norm:
\begin{align}
  e := \frac{1}{N} \sum_{m=1}^N |\phi^F(\x_m) - \phi^*(\x_m)|.\label{eq:err}
\end{align}
We draw $\x_m\in[0,1)^3$ from a uniform random distribution, randomize
charges under the constraints that $\sum q_m = 0$ and $\sum |q_m| =
1$. The minimal computational domain is $[0,1)^2\times[-w,1+w]$, but
as $w=w(P)$ we take $[0,1)^2\times[-1/2, 3/2]$ to avoid having a
PME-grid that depends on $P$. We take $N=50$, consider two cases:
$\xi=4$, $M=21$ and $\xi=12$, $M=50$. The convergence results in
Figure \ref{fig:se2p_pme_acc} support several conclusions: $(i)$
spectral accuracy as predicted by theory; $(ii)$ convergence rate
independent on Ewald parameter $\xi$; $(iii)$ oversampling the grid in
the $z$-direction by a factor three (or six as $L_z = 2L_{x,y}$,
i.e. $M_z = 2 s_f M = 6M$, depending on how you look at it) is
sufficient to get double precision accuracy.

\subsection{Computational overview} \label{sec:runtime_prof}

To give a sense of the practical characteristics of our method, we
give a brief overview of the run-time profile with our
implementation. We have previously discussed the need to oversample
the Fourier transform in the $z$-direction, and the question then
naturally arises if this incurs a significant cost. We note that the
$(x,y)$-grid is $M\times M$ and, by the error analysis presented, $M$
rarely needs to be bigger than 50. Hence, even in the case where we
oversample by a factor 6 -- to safely commit an quadrature error in
the Fourier integral transform on the order of machine accuracy, cf
Section \ref{sec:f_int} -- the total grid size is $6\times 50^3 = 750
000$ elements (stored in about 6MB). In the works cited, grid sizes of
up to $512^3$ are mentioned (at much more modest accuracies).

\begin{figure}
  \centering
  \includegraphics{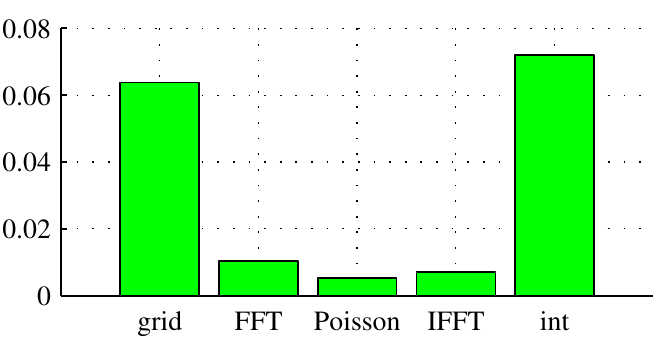}
  \includegraphics{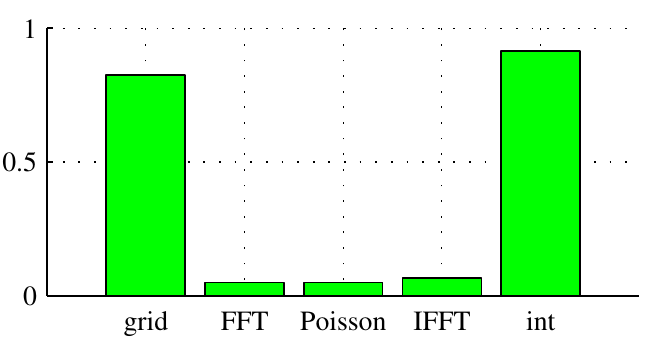}
  \caption{Runtime profile, i.e. time spent in different parts of fast
    algorithm, where ``grid'' refers to \eqref{eq:se2p_H_sum},
    ``Poisson'' refers to \eqref{eq:se2p_H_mult} and ``int'' refers to
    \eqref{eq:se2p_H_int}. Left: $N=10000$, $M=20$, $P=15$. Right:
    $N=10^5$, $M=40$, $P=17$. In both cases FFT oversampled by factor
    six in $z$-direction, $M_z = 6M$. Despite that, the transforms take
    a trivial amount of time to compute.}
  \label{fig:runtime_prof}
\end{figure}

As expected then, the computational burden in our method falls on the
gridding steps \eqref{eq:se2p_H_sum} and \eqref{eq:se2p_H_int}. We
illustrate this for two systems in Figure \ref{fig:runtime_prof}. Here
we let $\xi = 8$ and target an accuracy $\epsi \approx 10^{-10}$ (see
caption for further details). These are single-core results obtained
on an ordinary workstation computer (Intel Core2Duo E6600),
implementation in C. In Table \ref{tab:grid_cost} we give performance
numbers for the gridding step in terms of the support $P$.

\subsection{Scaling to large systems}
Computing the real-space sum \eqref{eq:rs_sum} has been ignored up to
this point, save for the remarks on parameter selection of Section
\ref{sec:param_choices}. In this regard we follow the conventional
line of thought -- computing \eqref{eq:rs_sum} has $\O(N)$ complexity
\emph{iff} each particle interacts with a \emph{fixed} number of
neighbors that lie within a radius $r_c$, as $N$ grows. This implies
either $(i)$ that the domain grows (so that $N/L^3$ constant, and
$r_c$ constant), or $(ii)$ that the interaction radius, $r_c$,
decreases, as $N$ grows. Regardless, the grid size, $M$, will grow.

Following the second approach, we return to the parameter estimates in
Section \ref{sec:param_choices} and note, by \eqref{eq:xi_est}, that
$\xi$ grows as $r_c$ becomes smaller. Consequently, and by
\eqref{eq:k_M_est}, the grid size $M$ will grow with $N$. Hence, the
complexity of our proposed spectrally accurate PME method is $\O(N\log
N)$. Note that before the $\O(N)$ complexity of the real space sum is
imposed, the calculations \eqref{eq:se2p_H_sum} -
\eqref{eq:se2p_H_int} have complexity $\O(NP^3) + \O(M^3 \log M^3) =
\O(N)$. As we suggest in the previous section, the constant in front
of the first term is quite a bit bigger than the constant in the FFT
part. Thus, the $\log N$ factor is not seen in practice.

\begin{figure}
  \begin{minipage}[t]{0.49\linewidth}
    \vspace{0pt}
    \centering
    \includegraphics{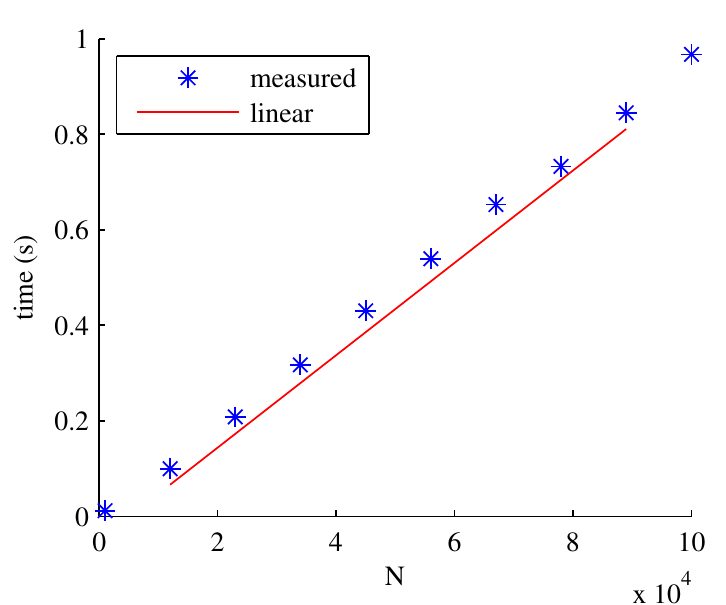}
  \end{minipage}
  \hspace{0.1cm}
  \begin{minipage}[t]{0.49\linewidth}
    \vspace{20pt}
    \centering 
    \begin{tabular}{r|lll}
      $N$ & $r_c$ & $\xi$ & $M$\\
      \hline
      1000 & 0.62 & 7.62 & 23 \\ 
      12000 & 0.27 & 17.90 & 55 \\ 
      23000 & 0.22 & 22.38 & 69 \\ 
      34000 & 0.19 & 25.60 & 79 \\ 
      45000 & 0.17 & 28.18 & 87 \\ 
      56000 & 0.16 & 30.38 & 94 \\ 
      67000 & 0.15 & 32.31 & 100 \\ 
      78000 & 0.15 & 34.04 & 105 \\ 
      89000 & 0.14 & 35.61 & 110 \\ 
      100000 & 0.13 & 37.07 & 115 \\ 
    \end{tabular}
  \end{minipage}
  \caption{Left: Run-time as a function of $N$, with
    $\epsi=10^{-9}$. Right: Parameters to scale up system at constant
    cost for real-space sum}
  \label{fig:scale_N}
\end{figure}

To verify this, and clarify the parameter selection process
(cf. Section \ref{sec:param_choices}), we give scaling results
(measured run-time to compute $\phi^F$) in Figure \ref{fig:scale_N},
including the parameter table (right). Here, the target accuracy is
$\epsi=10^{-9}$, we start with $N=1000$ and let $P=15$, and invoke the
estimates as described.

\begin{table}
  \centering
  \begin{tabular}{l|llllll}
    $P=$ & 3   &  7  &  11  &  15  &  19  &  23\\
    \hline
    time $[\mu s]$ & 0.73 & 1.67 & 3.36 & 7.48 & 13.57 & 23.49\\
  \end{tabular}

  \caption{Time for gridding (in microseconds per particle) 
    for different support $P$.}
  \label{tab:grid_cost}
\end{table}

\section{Summary and concluding remarks}

In our survey of methods to compute the sum of Coulomb potentials
\eqref{eq:pot_sum} we argue that methods for the 2P case are less
mature and consistent than their 3P cousins. Hence, the desire from
the applications community for an established tool for 2P
electrostatic calculations is to some extent unsatisfied.

We aim to close the gap between 2P and 3P Ewald methods by two
provisions. First, we derive the 2P Ewald sum \eqref{eq:2dp_ewald_sum}
using the established methodology of screening functions that follows
the 3P case closely (Section \ref{sec:ewald_deriv}). Secondly, we
derive a fast PME-type method for the 2P $\k$-space sum that fits well
in the established PME framework (Section \ref{sec:fast_fd_method}),
which we refer to as SE2P. These derivations were made possible by
representing functions on mixed periodicity ($x,y$ periodic, $z$
``free'') in frequency domain via a mixed Fourier transform
\eqref{eq:k_space_repr}, see Section \ref{sec:prelim}. This point of
view is natural and clarifies the relationship between 2P and 3P Ewald
methods to an extent that we do not believe has been previously
reported.

In light of this, we conclude that a fast PME-type method for 2P will
have to compute a mixed transform (a discrete Fourier transform in the
periodic variables, and an approximation to the continuous Fourier
integral transform in the free dimension). Efficiency constraints
suggest that the quadrature for the Fourier integral should be based
on the FFT. We studied this problem in Section \ref{sec:f_int}, and
point out that the accuracy of an FFT-based quadrature method will
depend on the regularity of the integrand.

Hence, a method using the SPME approach (using Cardinal B-splines to
represent regularized charges on the grid) will have to deal with
vastly reduced accuracy in the quadrature step of the Fourier
transform \emph{vis-a-vis} the approach that we suggest, which uses
Gaussians that are $C^\infty$ smooth. Whereas established PME methods
may be seen as adequate, in terms of accuracy, for the 3P case, our
analysis suggest that that may not be true in 2P (Section
\ref{sec:cbsp_fquad}). The SE2P method is similar in structure to the
work by Kawata et. al. \cite{Kawata2001a}, though it appears to offer
significant advantages.

The method we propose (Sections \ref{sec:fast_fd_method} to
\ref{sec:fgg}) for computing the 2P $\k$-space Ewald sum is spectrally
accurate, meaning that all errors decay exponentially (as we establish
theoretically in Section \ref{sec:error_analysis} and verify
numerically in Section \ref{sec:num_res}). More specifically, the
numerical errors present stem from two sources: truncation of the
underlying Ewald sum (Section \ref{sec:truncation}) and approximation
errors introduced by the fast method (Section \ref{sec:approx_err}). A
well established error estimate for the former is used to determine
the appropriate grid size, $M$. Our error analysis of the latter is
used to determine the number of points within the support of our
Gaussians, $P$. To our knowledge, this is the only fast Ewald
summation method that is spectrally accurate and the only one that
retains a decoupling of errors as discussed here -- \emph{a fortiori}
in 2P.

Moreover, the proposed method is efficient, capable of dealing with $N
\sim 10^6$ in a few seconds (Section \ref{sec:runtime_prof}). In
particular, we see that the grid sizes needed are very small -- so
small that the Fourier transforms are cheap to compute, even when
allowing for oversampling the $z$-dimension. The computational burden
falls more heavily on the gridding steps \eqref{eq:se2p_H_sum} and
\eqref{eq:se2p_H_int}. The Fast Gaussian Gridding approach (Section
\eqref{sec:fgg}) alleviates this to a large extent.

We believe that these properties -- accuracy, clear parameter
selection, efficiency, and closeness to 3P methods -- present a
compelling case for the proposed method for electrostatic
calculations in planar periodicity.

\section*{Acknowledgments}
A.K.T. is a Royal Swedish Academy of Sciences Research Fellow
supported by a grant from the Knut and Alice Wallenberg Foundation and
thankfully acknowledges this support.

\appendix
\section{Derivation of 2P Ewald sum, details}
\label{app:sing_deriv}

The derivation of $\phiz(z)$, the singularity contribution
\eqref{eq:fd_singular_sum}, is given here for completeness and
because it's illuminating in it's own right.

To make this clear, we first disregard the screening, $\gamma$,
i.e. consider
\begin{align*}
  -\Delta \phi(\x) = 4\pi\sumn \rho^n(\x), \quad \rho^n(\x) =
  \sum_{\p\in\Z^2} q_n \delta(\x-\x_n + \tilde{\p}),
\end{align*}
under an assumption of charge neutrality, $\sumn q_n \equiv 0$, and
the condition
\begin{align}
  \lim_{z\rightarrow\pm\infty} \phi(\x) = \pm \frac{2\pi}{L^2}
  \sumn q_nz_n. \label{eq:2pbc}
\end{align}
Provisionally, as in Section \ref{sec:ewald_deriv}, 
\begin{align*}
  \phi(\r,z) \sim \frac{2}{L^2} \sum_{\k} \int_\R
  \frac{1}{k^2+\kappa^2} \sumn q_n e^{i\k\cdot (\r-\r_n)} e^{i\kappa (z-z_n)}
  \d\kappa.
\end{align*}
The terms corresponding to $\k\neq 0$ are uncomplicated and can be
integrated,
\begin{align}
  \tilde{\phi}(\r,z) &:= \frac{2}{L^2} \sum_{\k\neq 0} \int_\R
  \frac{1}{k^2+\kappa^2} \sumn q_n e^{i\k\cdot(\r-\r_n)} e^{i\kappa (z-z_n)}
  \d\kappa \nonumber \\ &= \frac{2\pi}{L^2} \sum_{\k\neq 0} \sumn q_n
  \frac{1}{\|\k\|} e^{-\|\k\| |z-z_n|} e^{i\k\cdot(\r-\r_n)}.\label{eq:phi_full}
\end{align}
Note that $\lim_{z\rightarrow\pm\infty} \tilde{\phi} = 0$. Hence, we
seek a term $\phi^0(z)$ with the desired behavior \eqref{eq:2pbc} at
$z\rightarrow\pm\infty$. Then, $\phi(\x) = \tilde{\phi}(\x) +
\phi^0(z)$ will be a unique and well defined solution to the
2-periodic Poisson problem under consideration.

If we disregard the boundary condition \eqref{eq:2pbc}, it's evident
that $\phi$ is only determined up to a piecewise linear
function. Consider the adding the term $\phi^0(z)=b\sumn q_n
|z-z_n|$. Using charge neutrality, one finds that
\begin{align*}
  \lim_{x\rightarrow\pm\infty} \phi^0(z) = \mp b\sumn q_n z_n. 
\end{align*}

This solution, with $b=-\frac{2\pi}{L^2}$, can be found e.g. by
considering the one-dimensional Green's function for the $\k=0$
mode, see Genovese et. al. \cite{Genovese2007}. With this,
\begin{align*}
  \phi(\r,z) = \frac{2\pi}{L^2} \sum_{\k\neq 0} \sumn q_n
  \frac{1}{\|\k\|} e^{-\|\k\| |z-z_n|} e^{i\k\cdot(\r-\r_n)} -
  \frac{2\pi}{L^2} \sumn q_n|z-z_n|
\end{align*}
satisfies \eqref{eq:2pbc}.

There's a natural correspondence between $\tilde{\phi}$ and $\phi^F$
\eqref{eq:fd_sum}. However, the crucial point when the decomposition
\eqref{eq:decomp} enters is that some of the $\k=0$ mode will be
included into the real-space sum. Therefore, rather than taking
$\phiz$ equal to $\phi^0$, we subtract the real-space term (which is
most accessible as the difference between $\tilde{\phi}$ and
$\phi^F$),
\begin{align*}
  \phiz(z) &= \phi^0(z) - \lim_{\k\rightarrow 0} \left(
    \widehat{\tilde{\phi}}_\k - \widehat{\phi^F}_\k \right).
\end{align*}
We introduce 
\begin{align*}
  \lim_{\k\rightarrow 0} \left( \widehat{\phi^F}_\k -
    \widehat{\widehat{\phi}}_\k \right) = \frac{\pi}{L^2} \sumn q_n
  a(z-z_n),
\end{align*}
and compute
\begin{align*}
  a(z) &= \lim_{k\rightarrow 0} \frac{1}{k}\left( e^{kz}\erfc\left(\frac{k}{2\xi} + \xi z\right) + e^{-k z} \erfc\left(\frac{k}{2\xi} - \xi z\right) - 2 e^{-k |z|} \right )\\
  & = -\frac{2}{\sqrt{\pi}} \left( \frac{1}{\xi}e^{-\xi^2 z^2} +
    \sqrt{\pi}( -|z| + z \erf(\xi z)) \right).
\end{align*}
Finally,
\begin{align*}
  \phiz(z) &= \frac{\pi}{L^2} \sumn q_n a(z-z_n) - \frac{2\pi}{L^2} \sumn q_n |z-z_n|\\
  & = - \frac{2\sqrt{\pi}}{L^2} \sumn q_n \left(
    \frac{1}{\xi}e^{-\xi^2 (z-z_n)^2} + \sqrt{\pi} (z-z_n)
    \erf(\xi(z-z_n)) \right),
\end{align*}
as we set out to show. Using charge neutrality, the limits
\eqref{eq:2pbc} can be verified.

\bibliographystyle{plain}
\bibliography{se2p}{}

\end{document}